%% file: main.tex
\documentclass{llncs}
\usepackage[utf8]{inputenc}
\usepackage[T1]{fontenc}
\usepackage{amsmath}
\usepackage{amssymb}
\usepackage{latexsym}
\usepackage{cmll}
\usepackage{url}
\usepackage{bussproofs}

\usepackage{tikz-inet}

\usepackage{alltt}

\input local

\input notation
\input tikznotation

\title{Non-idempotent intersection types in logical form}
\author{Thomas Ehrhard}
\institute{IRIF, CNRS and Paris University, \email{ehrhard@irif.fr},
\url{www.irif.fr/~ehrhard/}}

\begin{document}
\maketitle

\label{firstpage}

\begin{abstract}
  Intersection types are an essential tool in the analysis of
  operational and denotational properties of lambda-terms and
  functional programs. Among them, non-idempotent intersection types
  provide precise quantitative information about the evaluation of
  terms and programs. However, unlike simple or second-order types,
  intersection types cannot be considered as a logical system because
  the application rule (or the intersection rule, depending on the
  presentation of the system) involves a condition expressing that the
  proofs of premises satisfy a very strong uniformity condition: the
  underlying lambda-terms must be the same. Using earlier work
  introducing an indexed version of Linear Logic, we show that
  non-idempotent typing can be given a logical form in a system where
  formulas represent hereditarily indexed families of intersection
  types.
\end{abstract}

\section*{Introduction}
Intersection types, introduced in the work of Coppo and
Dezani~\cite{CoppoDezani80,CoppoDezaniVenneri81} and developed since
then by many authors, are still a very active research topic.
As quite clearly explained in~\cite{Krivine93}, the Coppo and Dezani
intersection type system $D\Omega$ can be understood as a syntactic
presentation of the denotational interpretation of $\lambda$-terms in
the Engeler's model, which is a model of the pure $\lambda$-calculus
in the cartesian closed category of prime-algebraic complete lattices
and Scott continuous functions.

Intersection types can be considered as formulas of the propositional
calculus implication with $\Implies$ and conjunction $\wedge$ as
connectives. However, as pointed out by Hindley~\cite{Hindley84},
intersection types deduction rules depart drastically from the
standard logical rules of intuitionistic logic (and of any standard
logical system) by the fact that, in the $\wedge$-introduction rule,
it is assumed that the proofs of the two premises are typing of the
\emph{same} $\lambda$-term, which means that, in some sense made
precise by the typing system itself, they have the same
structure. Such requirements on \emph{proofs of} premises, and not
only on formulas proven in premises, are absent from standard
(intuitionistic or classical) logical systems where the proofs of
premises are completely independent from each other. Many authors have
addressed this issue, we refer to~\cite{LiquoriRonchi07} for a
discussion on several solutions which mainly focus on the design of
\emph{à la Church} presentations of intersection typing systems, thus
enriching $\lambda$-terms with additional structures. Among the most
recent and convincing contributions to this line of research we should
certainly mention~\cite{LiquoriStolze19}.

In our ``new''\footnote{Not so new since it dates back to
  our~\cite{BucciarelliEhrhard99}.} approach to this problem, we
change formulas instead of changing terms. It is based on a specific
model of Linear Logic (and thus of the $\lambda$-calculus): the
\emph{relational model}\footnote{It is fair to credit Girard for the
  introduction of this model since it appears at least implicitly
  in~\cite{Girard88c}. This model was probably known by many people in
  the Linear Logic community as a piece of folklore since the early
  1990's. It is presented formally
  in~\cite{BucciarelliEhrhard99}.}. In this quite simple\footnote{As
  one can guess, there is a price to pay for this apparent simplicity:
  it is the relative complexity of the definition of morphism
  composition and of duplication, which take carefully multiplicities
  into account, see Section~\ref{sec:relsem}.} and canonical
denotational model, types are interpreted as sets (without any
additional structure) and a closed term of type $\sigma$ is
interpreted as a subset of the interpretation of $\sigma$. It is quite
easy to define, in this semantic framework, analogues of the usual
models of the pure $\lambda$-calculus such as Scott's $D_\infty$ or
Engeler's model, which in some sense are simpler than the original
ones since the sets interpreting types need not to be pre-ordered. As
explained in the work of De~Carvalho~\cite{DeCarvalho09,DeCarvalho18},
the intersection type counterpart of this semantics is a typing system
where ``intersection'' is non-idempotent (in sharp contrast with the
original systems introduced by Coppo and Dezani), sometimes called
\emph{system $R$}. Notice that the precise connection between the
idempotent and non-idempotent approaches is analyzed
in~\cite{Ehrhard11b}, in a quite general Linear Logic setting by means
of an extensional collapse\footnote{This shows that, when one wants to
  ``forget multiplicities'' in the relational model, one needs to
  equip the sets interpreting types with a preorder relation.}.

In order to explain our approach, we restrict first to simple types,
interpreted as follows in the relational model: a basic type $\alpha$
is interpreted as a given set $\Tsem\alpha$ and the type
$\Timpl\sigma\tau$ is interpreted as the set
$\Mfin{\Tsem\sigma}\times\Tsem\tau$ (where $\Mfin E$ is the set of
finite multisets of elements of $E$). Remember indeed that
intersection types can be considered as a syntactic presentation of
denotational semantics, so it makes sense to define intersection types
relative to simple types (in the spirit of~\cite{FreemanPfenning91})
as we do in Section~\ref{sec:simple-types}: an intersection type
relative to the base type $\alpha$ is an element of $\Tsem\alpha$ and
an intersection type relative to $\Timpl\sigma\tau$ is a pair
$(\Mfin{\List a1n},b)$ where the $a_i$s are intersection types
relative to $\sigma$ and $b$ is an intersection type relative to
$\tau$; with more usual notations\footnote{That we prefer not to use
  for avoiding confusions between these two levels of typing.}
$(\Mfin{\List a1n},b)$ would be written
$(a_1\wedge\cdots\wedge a_n)\to b$. Then, given a type $\sigma$, the
main idea consists in representing an indexed family of elements of
$\Tsem\sigma$ as a formula of a new logical system. If
$\sigma=\Timplp\phi\psi$ then the family can be written\footnote{We
  use $\Mset\cdots$ for denoting multisets much as one uses
  $\Eset\cdots$ for denoting sets, the only difference is that
  multiplicities are taken into account.}
$(\Mset{a_k\St k\in K\text{ and }u(k)=j},b_j)_{j\in J}$ where $J$ and
$K$ are indexing sets, $u:K\to J$ is a function such that
$\Funinv f(\Eset j)$ is finite for all $j\in J$, $(b_j)_{j\in J}$ is a
family of elements of $\Tsem\psi$ (represented by a formula $B$) and
$(a_k)_{k\in K}$ is a family of elements of $\Tsem\phi$ (represented
by a formula $A$): in that case we introduce the implicative formula
$\Tfunp AuB$ to represent the family
$(\Mset{a_k\St k\in K\text{ and }u(k)=j},b_j)_{j\in J}$. It is clear
that a family of simple types has generally infinitely many
representations as such formulas; this huge redundancy makes it
possible to establish a tight link between inhabitation of
intersection types with provability (in an indexed version $\LJ(I)$ of
intuitionistic logic) of formulas representing them. Such a
correspondence is exhibited in Section~\ref{sec:simple-types} in the
simply typed setting and the idea is quite simple:
\begin{quote}
  given a type $\sigma$, a family $(a_j)_{j\in J}$ of elements of
  $\Tsem\sigma$, and a closed $\lambda$-term of
  type $\sigma$, it is equivalent to say that $\Tseq{}M{a_j}$ holds
  for all $j$ and to say that some (and actually any) formula $A$
  representing $(a_j)_{j\in J}$ has an $\LJ(I)$ proof\footnote{Any
    such proof can be stripped from its indexing data giving rise to a
    proof of $\sigma$ in intuitionistic logic.} whose underlying
  $\lambda$-term is $M$.
\end{quote}

In Section~\ref{sec:D-infty-relational} we extend this approach to the
untyped $\lambda$-calculus taking as underlying model of the pure
$\lambda$-calculus the aforementioned relational version of Scott's
$D_\infty$ that we denote as $\Dinfr$: it is the least set which
contains all $\Nat$-indexed sequences $a$ of finite multisets of
elements $\Dinfr$ such that $a_n$ is the empty multiset for almost all
$a$. We define an adapted version of $\LJ(I)$ and establish a similar
correspondence, with some slight modifications due to the
specificities of $\Dinfr$.

\section{Notations and preliminary definitions}\label{sec:notations}

If $E$ is a set, a \emph{finite multiset of elements of $E$} is a
function $m:E\to\Nat$ such that the set $\Eset{a\in E\St m(a)\not=0}$
(called the \emph{domain} of $m$) is finite. The cardinal of such a
multiset $m$ is $\Card m=\sum_{a\in E}m(a)$. We use $+$ for the
obvious addition operation on multisets, and if $\List a1n$ are
elements of $E$, we use $\Mset{\List a1n}$ for the corresponding
multiset (taking multiplicities into account); for instance
$\Mset{0,1,0,2,1}$ is the multiset $m$ of elements of $\Nat$ such that
$m(0)=2$, $m(1)=2$, $m(2)=1$ and $m(i)=0$ for $i>2$.  If
$(a_i)_{i\in I}$ is a family of elements of $E$ and if $J$ is a finite
subset of $I$, we use $\Mset{a_i\St i\in J}$ for the multiset of
elements of $E$ which maps $a\in E$ to the number of elements $i\in J$
such that $a_i=a$ (which is finite since $J$ is).
We use $\Mfin E$ for the set of finite multisets of elements of $E$.

We use $+$ to denote set union when we we want to stress the fact that
the invloved sets are disjoint.  A function $u:J\to K$ is \emph{almost
  injective} if $\Card{\Funinv f\Eset k}$ is finite for each $k\in K$
(so the inverse image of any finite subset of $K$ under $u$ is
finite).  If $s=(\List a1n)$ is a sequence of elements of $E$ and
$i\in\Eset{1,\dots,n}$, we use $\Seqext si$ for the sequence
$(a_1,\dots,a_{i-1},a_{i+1},\dots,a_n)$.  Given sets $E$ and $F$, we
use $F^E$ for the set of function from $E$ to $F$. The elements of
$F^E$ are sometimes considered as functions $u$ (with a functional
notation $u(e)$ for application) and sometimes as indexed families $a$
(with index notations $a_e$ for application) especially when $E$ is
countable. The choice between these options should be clear from the
context.

If $i\in\Eset{1,\dots,n}$ and $j\in\Eset{1,\dots,n-1}$, we define
$\Csucc ji\in\Eset{1,\dots,n}$ as follows: $\Csucc ji=j$ if
$j<i$ and $\Csucc ji=j+1$ if $j\geq i$.

\section{The relational model of the $\lambda$-calculus}
\label{sec:relsem}

Let $\RELK$ the category whose objects are sets\footnote{We can restrict
  to countable sets.} and $\RELK(X,Y)=\Part{\Mfin X\times Y}$ with
$\Id_X=\Eset{(\Mset a,a)\St a\in X}$ and composition of
$s\in\RELK(X,Y)$ and $t\in\RELK(Y,Z)$ given by
\begin{align*}
  t\Comp s&=
            \Eset{(m_1+\cdots+m_k,c)\St\\
  &\hspace{1cm}\exists \List b1k\in Y\
  (\Mset{\List b1k},c)\in t\text{ and }\forall j\,(m_j,b_j)\in s}\,.
\end{align*}
It is easily checked that this composition law is associative and that
$\Id$ is neutral for composition\footnote{This results from the fact
  that $\RELK$ arises as the Kleisli category of the \LL{} model of
  sets and relations, see~\cite{BucciarelliEhrhard99} for
  instance.}. This category has all countable products: let
$(X_j)_{j\in J}$ be a countable family of sets, their product is
$X=\Bwith_{j\in J}X_j=\Union_{j\in J}\Eset j\times X_j$ and
projections $(\Proj j)_{j\in J}$ given by
$\Proj j=\Eset{(\Mset{(j,a)},a)\St a\in X_j}\in\RELK(X,X_j)$ and if
$(s_j)_{j\in J}$ is a family of morphisms $s_j\in\RELK(Y,X_j)$ then
their tupling is
$\Tuple{s_j}_{j\in J}=\Eset{(\Mset a,(j,b)))\St j\in J\text{ and
  }(\Mset a,b)\in s_j}\in\RELK(Y,X)$.

The category $\RELK$ is cartesian closed with object of morphisms from
$X$ to $Y$ the set $(\Simpl XY)=\Mfin X\times Y$ and evaluation
morphism $\Ev\in\RELK(\With{(\Simpl XY)}{X},Y)$ is given by
\(
  \Ev=\Eset{(\Mset{(1,\Mset{\List a1k},b),(2,a_1),\dots,(2,a_k)},b)
  \St \List a1k\in X\text{ and }b\in Y}
\).
The transpose (or curryfication) of $s\in\RELK(\With ZX,Y)$ is
$\Curry s\in\RELK(Z,\Simpl XY)$ given by
\(
  \Curry s =\Eset{(\Mset{\List c1n},(\Mset{\List a1k},b))\St
  (\Mset{(1,c_1),\dots,(1,c_n),(2,a_1),\dots,(2,a_k)},c)\in s}
\).

\paragraph*{Relational $D_\infty$.}

Let $\Dinfr$ be the least set such that $(m_0,m_1,\dots)\in\Dinfr$ as
soon as $m_0,m_1\dots$ are finite multisets of elements of $\Dinfr$
which are almost all equal to $\Mset{}$. Notice in particular that
$\Estack=(\Mset{},\Mset{},\dots)\in\Dinfr$ and satisfies
$\Estack=(\Mset{},\Estack)$. By construction we have
$\Dinfr=\Mfin \Dinfr\times\Dinfr$, that is
$\Dinfr=(\Simpl\Dinfr\Dinfr)$ and hence $\Dinfr$ is a model of the
pure $\lambda$-calculus in $\RELK$ which also satisfies the
$\eta$-rule.  See~\cite{BreuvartManzonettoRuoppolo18} for general facts on
this kind of model.

\section{The simply typed case}
\label{sec:simple-types}

We assume to be given a set of type atoms $\alpha,\beta,\dots$ and of
variables $x,y,\dots$; types and terms are given as usual by
\(
\sigma,\tau,\dots \Bnfeq \alpha \Bnfor \Timpl\sigma\tau
\)
and
\(
  M,N,\dots \Bnfeq x \Bnfor \App MN \Bnfor \Abst x\sigma N
\).

With any type atom we associate a set $\Tsem\alpha$. This
interpretation is extended to all types by
$\Tsem{\Timpl\sigma\tau}=\Simpl{\Tsem\sigma}{\Tsem\tau}=
\Mfin{\Tsem\sigma}\times\Tsem\tau$.  The relational semantics of this
$\lambda$-calculus can be described as a non-idempotent intersection
type system, with judgments of shape
\(
  \Tseq{x_1:m_1:\sigma_1,\dots,x_n:m_n:\sigma_n}{M}{a:\sigma}
\)
where the $x_i$'s are pairwise distinct variables, $M$ is a term,
$a\in\Tsem\sigma$ and $m_i\in\Mfin{\Tsem{\sigma_i}}$ for each
$i$. Here are the typing rules:
\begin{center}
  \AxiomC{$j\not=i\Implies m_j=\Mset{}$ and $m_i=\Mset a$}
  \UnaryInfC{$\Tseq{\Family{x_i:m_i:\sigma_i}i1n}
    {x_i}{a:\sigma}$}
  \DisplayProof
  \quad
  \AxiomC{$\Tseq{\Phi,x:m:\sigma}{M}{b:\tau}$}
  \UnaryInfC{$\Tseq{\Phi}
    {\Abst{x}{\sigma}M}{(m,b):\Timpl{\sigma}\tau}$}
  \DisplayProof
\end{center}

\begin{center}
  \AxiomC{$\Tseq{\Phi}{M}
    {(\Mset{\List a1k},b):\Timpl\sigma\tau}$}
  \AxiomC{$(\Tseq{\Phi_l}{N}
    {a_l:\sigma})_{l=1}^k$}
  \BinaryInfC{$\Tseq{\Psi}
    {\App MN}{b:\tau}$}
  \DisplayProof  
\end{center}
where $\Phi=\Family{x_i:m_i:\sigma_i}i1n$,
$\Phi_l=\Family{x_i:m^l_i:\sigma_i}i1n$ for $l=1,\dots,k$ and
\( \Psi=\Family{x_i:m_i+\sum_{l=1}^k m_i^l:\sigma_i}i1n \).

\subsection{Why do we need another system?}

The trouble with this deduction system is that it cannot be considered
as the term decorated version of an underlying ``logical system for
intersection types'' allowing to prove sequents of shape
\(
  \Jseq{m_1:\sigma_1,\dots,m_n:\sigma_n}{a:\sigma}
\)
(where non-idempotent intersection types $m_i$ and $a$ are considered
as logical formulas, the ordinary types $\sigma_i$ playing the role of
``kinds'') because, in the application rule above, it is required that
all the proofs of the $k$ right hand side premises have the same shape
given by the $\lambda$-term $N$. We propose now a ``logical system''
derived from~\cite{BucciarelliEhrhard99} which, in some sense, solves
this issue. The main idea is quite simple and relies on three
principles:
(1) replace \emph{hereditarily} multisets with indexed families in
  intersection types,
(2) instead of proving single types, prove indexed families of
  hereditarily indexed types
and (3) represent syntactically such families (of hereditarily indexed
  types) as formulas of a new system of \emph{indexed logic}.


\subsection{Minimal $\LJ(I)$}

We define now the syntax of indexed formulas. Assume to be given an
infinite countable set $I$ of indices (one can take $I=\Nat$, but we
use no specific properties or structures on natural numbers). Then we
define indexed types $A$; with each such type we associate an
underlying type $\Tunder A$, and a family
$\Tfam A\in\Tsem{\Tunder A}^{\Tdom A}$.  These formulas are given by
the following inductive definition:
\begin{itemize}
\item if $J\subseteq I$ and $f:J\to\Tsem\alpha$ is a function then
  $\Tcst\alpha f$ is a formula with $\Tunder{\Tcst\alpha f}=\alpha$,
  $\Tdom{\Tcst\alpha f}=J$ and $\Tfam{\Tcst\alpha f}=f$
\item and if $A$ and $B$ are formulas and $u:\Tdom A\to\Tdom B$ is
  almost injective then $\Tfun AuB$ is a formula with
  $\Tunder{\Tfun AuB}=\Timpl{\Tunder A}{\Tunder B}$,
  $\Tdom{\Tfun AuB}=\Tdom B$ and, for $k\in\Tdom B$,
  $\Tfam{\Tfun AuB}_k=(\Mset{\Tfam A_j\St j\in\Tdom A\text{ and
    }u(j)=k},\Tfam B_j)$.
\end{itemize}

\begin{proposition}\label{prop:family-rep}
  Let $\sigma$ be a type, $J$ be a subset of $I$ and
  $f\in\Tsem\sigma^J$. There is a formula $A$ such that
  $\Tunder A=\sigma$, $\Tdom A=J$ and $\Tfam A=f$ (actually,
  there are infinitely many such $A$'s as soon as $\sigma$ is not an
  atom and $J\not=\emptyset$).
\end{proposition}
\Beginproof
The proof is by induction on $\sigma$. If $\sigma$ is an atom $\alpha$
then we take $A=\Tcst\alpha f$. Assume that $\sigma=(\Timpl\rho\tau)$
so that $f(j)=(m_j,b_j)$ with $m_j\in\Mfin{\Tsem\rho}$ and
$b_j\in\Tsem\tau$. Since each $m_j$ is finite and $I$ is infinite, we
can find a family $(K_j)_{j\in J}$ of pairwise disjoint finite subsets
of $I$ such that $\Card{K_j}=\Card{m_j}$. Let $K=\Union_{j\in J}K_j$,
there is a function $g:K\to\Tsem\rho$ such that
$m_j=\Mset{g(k)\St k\in K_j}$ for each $j\in J$ (choose first an
enumeration $g_j:K_j\to\Tsem\rho$ of $m_j$ for each $j$ and then
define $g(k)=g_j(k)$ where $j$ is the unique element of $J$ such that
$k\in K_j$). Let $u:K\to J$ be the unique function such that
$k\in K_{u(k)}$ for all $k\in K$; since each $K_j$ is finite, this
function $u$ is almost injective. By inductive hypothesis there is a
formula $A$ such that $\Tunder A=\rho$, $\Tdom A=K$ and $\Tfam A=g$,
and there is a formula $B$ such that $\Tunder B=\tau$, $\Tdom B=J$ and
$\Tfam B=(b_j)_{j\in J}$. Then the formula $\Tfun AuB$ is well formed
(since $u$ is an almost injective function $\Tdom A=K\to\Tdom B=J$)
and satisfies $\Tunder{\Tfun AuB}=\sigma$, $\Tdom{\Tfun AuB}=J$ and
$\Tfam{\Tfun AuB}=f$ as contended.
\Endproof

As a consequence, for any type $\sigma$ and any element $a$
of $\Tsem\sigma$ (so $a$ is a non-idempotent intersection type of kind
$\sigma$), one can find a formula $A$ such that $\Tunder A=\sigma$,
$\Tdom A=\Eset j$ (where $j$ is an arbitrary element of $J$) and
$\Tfam A_j=a$. In other word, any intersection type can be represented
as a formula (in infinitely many different ways in general of course,
but up to renaming of indices, that is, up to
``heriditary $\alpha$-equivalence'', this representation is unique).

For any formula $A$ and $J\subseteq I$, we define a formula
$\Trestr AJ$ such that $\Tunder{\Trestr AJ}=\Tunder A$,
$\Tdom{\Trestr AJ}=\Tdom A\cap J$ and
$\Tfam{\Trestr AJ}=\Tfam A\restriction_J$. The definition is by
induction on $A$.

\begin{itemize}
\item $\Trestr{\Tcst\alpha f}J=\Tcst\alpha{f\restriction_J}$
\item $\Trestr{(\Tfun AuB)}J=\Tfunp{\Trestr AK}v{\Trestr BJ}$ where
  $K=\Funinv v(\Tdom B\cap J)$ and $v=u\restriction_K$.
\end{itemize}

Let $u:\Tdom A\to J$ be a \emph{bijection}, we define a formula
$\Treloc uA$ such that $\Tunder{\Treloc uA}=\Tunder A$,
$\Tdom{\Treloc uA}=u(\Tdom A)$ and
$\Tfam{\Treloc uA}_j=\Tfam A_{\Funinv u(j)}$. The definition is by
induction on $A$:
\begin{itemize}
\item $\Treloc u{\Tcst\alpha f}=\Tcst\alpha{f\Comp\Funinv u}$
\item $\Treloc u{\Tfun AvB}=(\Tfun{A}{u\Comp v}{\Treloc uB})$.
\end{itemize}

Using these two auxiliary notions, we can give a set of three deduction
rules for a minimal natural deduction allowing to prove formulas in
this indexed intuitionistic logic. This logical system allows to
derive sequents which are of shape
\begin{align}\label{eq:indexed-logic-sequent}
  \Jseq{\Threloc {A_1}{u_1},\dots,\Threloc {A_n}{u_n}}{B}
\end{align}
where for each $i=1,\dots,n$, the function $u_i:\Tdom{A_i}\to\Tdom B$
is almost injective. Notice that the expressions $\Threloc{A_i}{u_i}$
are not formulas; this
construction $\Threloc{A}u$ is part of the syntax of sequents, just as
the ``$,$'' separating these pseudo-formulas. Given a formula $A$ and
$u:\Tdom A\to J$ almost injective, it is nevertheless convenient to define
$\Tfam{\Threloc Au}\in\Mfin{\Tsem{\Tunder A}}^J$ by
$\Tfam{\Threloc Au}_j=\Mset{\Tfam A_k\St u(k)=j}$. In particular, when
$u$ is a bijection, $\Tfam{\Threloc Au}_j=\Mset{\Tfam{A}_{\Funinv u(j)}}$.

The crucial point here is that such a
sequent~\Eqref{eq:indexed-logic-sequent} involves no $\lambda$-term.

The main difference between the original system $\LL(I)$
of~\cite{BucciarelliEhrhard99} and the present system is the way axioms
are dealt with. In $\LL(I)$ there is no explicit identity axiom and
only ``atomic axioms'' restricted to the basic constants of $\LL$;
indeed it is well-known that in $\LL$ all identity axioms can be
$\eta$-expanded, leading to proofs using only such atomic axioms. In
the $\lambda$-calculus, and especially in the untyped
$\lambda$-calculus we want to deal with in next sections, such
$\eta$-expansions are hard to handle so we prefer to use explicit
identity axioms.

The axiom is
%
\begin{center}
  \AxiomC{$j\not=i\Implies\Tdom{A_j}=\emptyset$ and $u_i$ is a bijection}
  \UnaryInfC{$\Jseq{\Threloc {A_1}{u_1},\dots,\Threloc {A_n}{u_n}}
    {\Treloc{u_i}{A_i}}$}
  \DisplayProof
\end{center}
so that for $j\not=i$, the function $u_j$ is empty. A special case is
\begin{center}
  \AxiomC{$j\not=i\Implies\Tdom{A_j}=\emptyset$ and $u_i$ is the
    identity function}
  \UnaryInfC{$\Jseq{\Threloc {A_1}{u_1},\dots,\Threloc {A_n}{u_n}}
    {A_i}$}
  \DisplayProof
\end{center}
which may look more familiar, but the general axiom rule, allowing to
``delocalize'' the proven formula $A_i$ by an arbitrary bijection
$u_i$, is required as we shall see%
.
The $\Implies$ introduction rule is quite simple
\begin{center}
  \AxiomC{$\Jseq{\Threloc {A_1}{u_1},\dots,\Threloc {A_n}{u_n},\Threloc Au}
    {B}$}
  \UnaryInfC{$\Jseq{\Threloc {A_1}{u_1},\dots,\Threloc {A_n}{u_n}}
    {\Tfun{A}{u}B}$}
  \DisplayProof
\end{center}

Last the $\Implies$ elimination rule is more complicated (from a
Linear Logic point of view, this is due to the fact that it combines
$3$ $\LL$ logical rules: $\Limpl{}{}$ elimination, contraction and
promotion). We have the deduction

\begin{center}
  \AxiomC{$\Jseq{\Threloc {C_1}{u_1},\dots,\Threloc {C_n}{u_n}}
    {\Tfun AuB}$}
  \AxiomC{$\Jseq{\Threloc {D_1}{v_1},\dots,\Threloc {D_n}{v_n}}
    {A}$}
  \BinaryInfC{$\Jseq{\Threloc {E_1}{w_1},\dots,\Threloc {E_n}{w_n}}
    {B}$}
  \DisplayProof
\end{center}
under the following conditions, to be satisfied by the involved
formulas and functions: for each $i=1,\dots,n$ one has
$\Tdom{C_i}\cap\Tdom{D_i}=\emptyset$,
$\Tdom{E_i}=\Tdom{C_i}+\Tdom{D_i}$, $C_i=\Trestr{E_i}{\Tdom{C_i}}$,
$D_i=\Trestr{E_i}{\Tdom{D_i}}$, $\Funrestr{w_i}{\Tdom{C_i}}=u_i$, and
$\Funrestr{w_i}{\Tdom{D_i}}=u\Comp v_i$.

Let $\pi$ be a deduction tree of the sequent
$\Jseq{\Threloc {A_1}{u_1},\dots,\Threloc {A_n}{u_n}} {B}$ in this
system. By dropping all index information we obtain a derivation tree
$\Tunder\pi$ of $\Jseq{\Tunder{A_1},\dots,\Tunder{A_n}}{\Tunder B}$,
and, upon choosing a sequence $\Vect x$ of $n$ pairwise distinct
variables, we can associate with this derivation tree a simply typed
$\lambda$-term $\Tunderl\pi{\Vect x}$ which satisfies
\(
  \Tseq{x_1:\Tunder{A_1},\dots,x_n:\Tunder{A_n}}
  {\Tunderl\pi{\Vect x}}{\Tunder B}
\).

\subsection{Basic properties of $\LJ(I)$}
We prove some basic properties of this logical system. This is also
the opportunity to get some acquaintance with it. Notice that in many
places we drop the type annotations of variables in $\lambda$-terms,
first because they are easy to recover, and second because the very
same results and proofs are also valid in the untyped setting of
Section~\ref{sec:D-infty-relational}.

\begin{lemma}[Weakening]\label{lemma:weakening}
  Assume that $\Jseq\Phi A$ is provable by a proof $\pi$ and let $B$
  be a formula such that $\Tdom B=\emptyset$. Then $\Jseq{\Phi'}{A}$
  is provable by a proof $\pi'$, where $\Phi'$ is obtained by
  inserting $\Threloc B{\Fempty{\Tdom A}}$ at any place in
  $\Phi$. Moreover $\Tunderl\pi{\Vect x}=\Tunderl{\pi'}{\Vect{x'}}$
  (where $\Vect{x'}$ is obtained from $\Vect x$ by inserting a dummy
  variable at the same place).
\end{lemma}
The proof is an easy induction on the proof of $\Jseq\Phi A$.

\begin{lemma}[Relocation]\label{lemma:reloc}
  Let $\pi$ be a proof of $\Jseq{(\Threloc{A_i}{u_i})_{i=1}^n}{A}$ let
  $u:\Tdom A\to J$ be a bijection, there is a proof $\pi'$ of
  $\Jseq{(\Threloc{A_i}{u\Comp u_i})_{i=1}^n}{\Treloc uA}$ such that
  $\Tunderl{\pi'}{\Vect x}=\Tunderl{\pi}{\Vect x}$.
\end{lemma}
The proof is a straightforward induction on $\pi$.

\begin{lemma}[Restriction]\label{lemma:restriction}
  Let $\pi$ be a proof of $\Jseq{\Family{\Threloc{A_i}{u_i}}i1n}A$ and let
  $J\subseteq\Tdom A$. For $i=1,\dots,n$, let
  $K_i=\Funinv{u_i}(J)\subseteq\Tdom{A_i}$ and
  $u'_i=\Frestr{u_i}{K_i}:K_i\to J$. Then the sequent
  $\Jseq{\Family{\Threloc{(\Trestr{A_i}{K_i})}{u'_i}}i1n}{\Trestr AJ}$
  has a proof $\pi'$ such that
  $\Tunderl{\pi'}{\Vect x}=\Tunderl{\pi}{\Vect x}$.
\end{lemma}
\Beginproof
By induction on $\pi$. Assume that $\pi$ consists of an axiom
$\Jseq{\Family{\Threloc{A_j}{u_j}}j1n}{\Treloc{u_i}{A_i}}$ with
$\Tdom{A_j}=\emptyset$ if $j\not=0$, and $u_i$ a bijection. With the
notations of the lemma, $K_j=\emptyset$ for $j\not=i$ and $u'_i$ is a
bijection $K_i\to J$. Moreover
$\Treloc{u'_i}{\Trestr{A_i}{K_i}}=\Trestr{\Treloc{u_i}{A_i}}{J}$ so
that
$\Jseq{\Family{\Threloc{(\Trestr{A_i}{K_i})}{u'_i}}i1n}{\Trestr AJ}$ is
obtained by an axiom $\pi'$ with
$\Tunderl{\pi'}{\Vect x}=x_i=\Tunderl{\pi}{\Vect x}$.

Assume that $\pi$ ends with a $\Implies$-introduction rule:
\begin{center}
  \AxiomC{$\rho$}
  \noLine
  \UnaryInfC{$\Jseq{\Family{\Threloc{A_i}{u_i}}i1{n+1}}{B}$}
  \UnaryInfC{$\Jseq{\Family{\Threloc{A_i}{u_i}}i1{n}}
    {\Tfun {A_{n+1}}{u_{n+1}}B}$}
  \DisplayProof
\end{center}
with $A=\Tfunp {A_{n+1}}{u_{n+1}}B$, and we have
$\Tunderl{\pi}{\Vect x}=\Abs{x_{n+1}}{\Tunderl{\rho}{\Vect
    x,x_{n+1}}}$. With the notations of the lemma we have
$\Trestr AJ=\Tfunp{\Trestr{A_{n+1}}{K_{n+1}}}{u'_{n+1}}{\Trestr
  BJ}$. By inductive hypothesis there is a proof $\rho'$ of
$\Jseq{\Family{\Threloc{\Trestr{A_i}{K_i}}{u'_i}}i1{n+1}}{\Trestr BJ}$
such that
$\Tunderl{\rho'}{\Vect x,x_{n+1}}=\Tunderl{\rho}{\Vect x,x_{n+1}}$ and
hence we have a proof $\pi'$ of
$\Jseq{\Family{\Threloc{\Trestr{A_i}{K_i}}{u'_i}}i1{n}}{\Trestr AJ}$
with
$\Tunderl{\pi'}{\Vect x}=\Abs{x_{n+1}}{\Tunderl{\rho'}{\Vect
    x,x_{n+1}}}=\Tunderl{\pi}{\Vect x}$ as contended.

Assume last that $\pi$ ends with a $\Implies$-elimination rule:
\begin{center}
  \AxiomC{$\mu$}
  \noLine
  \UnaryInfC{$\Jseq{\Family{\Threloc{B_i}{v_i}}i1n}{\Tfun BvA}$}
  \AxiomC{$\rho$}
  \noLine
  \UnaryInfC{$\Jseq{\Family{\Threloc{C_i}{w_i}}i1n}{B}$}
  \BinaryInfC{$\Jseq{\Family{\Threloc{A_i}{u_i}}i1n}{A}$}
  \DisplayProof
\end{center}
with $\Tdom{A_i}=\Tdom{B_i}+\Tdom{C_i}$,
$B_i=\Frestr{A_i}{\Tdom{B_i}}$ and $C_i=\Frestr{A_i}{\Tdom{C_i}}$,
$\Frestr{u_i}{\Tdom{B_i}}=v_i$ and
$\Frestr{u_i}{\Tdom{C_i}}=v\Comp w_i$ for $i=1,\dots,n$, and of course
$\Tunderl\pi{\Vect x}=\App{\Tunderl\mu{\Vect
    x}}{\Tunderl\rho{\Vect x}}$.  Let
$L=\Funinv v(J)\subseteq\Tdom B$. Let $L_i=\Funinv{v_i}(J)$ and
$R_i=\Funinv{w_i}{(L)}$ for $i=1,\dots,n$ (we also set
$v'_i=\Frestr{v_i}{L_i}$, $w'_i=\Frestr{w_i}{R_i}$ and
$v'=\Frestr vL$). By inductive hypothesis, we have a proof $\mu'$
of
$\Jseq{\Family{\Threloc{\Trestr{B_i}{L_i}}{v'_i}}i1n} {\Tfun{\Trestr
    BL}{v'}{\Trestr AJ}}$ such that
$\Tunderl{\mu'}{\Vect x}=\Tunderl\mu{\Vect x}$ and a proof
$\rho'$ of
$\Jseq{\Family{\Threloc{\Trestr{C_i}{R_i}}{w'_i}}i1n}{\Trestr BL}$
such that $\Tunderl{\rho'}{\Vect x}=\Tunderl\rho{\Vect x}$. Now,
setting $K_i=\Funinv{u_i}(K)$, observe that
\begin{itemize}
\item $\Tdom{B_i}\cap K_i=L_i=\Tdom{\Trestr{B_i}{L_i}}$
  and $\Frestr{u_i}{L_i}=v'_i$ since $\Frestr{u_i}{\Tdom{B_i}}=v_i$
\item $\Tdom{C_i}\cap K_i=R_i=\Tdom{C_i}\cap\Funinv{w_i}(L)$
  since $\Frestr{u_i}{\Tdom{C_i}}=v\Comp w_i$ and $L=\Funinv v(J)$,
  hence $\Tdom{C_i}\cap K_i=\Tdom{\Trestr{C_i}{R_i}}$, and
  also $\Frestr{u_i}{L_i}=v'\Comp w'_i$.
\end{itemize}
It follows that $\Tdom{\Trestr{A_i}{K_i}}=L_i+R_i$, and, setting
$u'_i=\Frestr{u_i}{K_i}$, we have $\Frestr{u'_i}{L_i}=v'_i$ and
$\Frestr{u'_i}{R_i}=v'\Comp w'_i$.  Hence we have a proof $\pi'$ of
$\Jseq{\Family{\Threloc{\Trestr{A_i}{K_i}}{u'_i}}i1n}{\Trestr AJ}$
such that
$\Tunderl{\pi'}{\Vect x}=\App{\Tunderl{\mu'}{\Vect
    x}}{\Tunderl{\rho'}{\Vect x}}=\App{\Tunderl{\mu}{\Vect
    x}}{\Tunderl{\rho}{\Vect x}}=\Tunderl\pi{\Vect x}$ as contended.
\Endproof

Though substitution lemmas are usually trivial, the $\LJ(I)$
substitution lemma requires some care in its statement and
proof\footnote{We use notations introduced in
  Section~\ref{sec:notations}.}.
\begin{lemma}[Substitution]\label{lemma:substitution}
  Assume that $\Jseq{(\Threloc{A_j}{u_j})_{j=1}^n}{A}$ with a proof
  $\mu$ and that, for some $i=1,\dots,n$,
  $\Jseq{(\Threloc{B_j}{v_j})_{j=1}^{n-1}}{A_i}$ with a proof
  $\rho$. Then there is a proof $\pi$ of
  $\Jseq{(\Threloc{C_j}{w_j})_{j=1}^{n-1}}{A}$ such that
  $\Tunderl\pi{\Seqext{\Vect x}i}=\Subst{\Tunderl\mu{\Vect
      x}}{\Tunderl\rho{\Seqext{\Vect x}i}}{x_i}$ as soon as for each
  $j=1,\dots,n-1$, $\Tdom{C_j}=\Tdom{A_{\Csucc ji}}+\Tdom{B_j}$
  for each $j=1,\dots,n-1$ with:
  \begin{itemize}
  \item $\Trestr{C_j}{\Tdom{A_{\Csucc ji}}}=A_{\Csucc ji}$
    and $\Frestr{w_j}{\Tdom{A_{\Csucc ji}}}=u_{\Csucc ji}$
  \item $\Trestr{C_j}{\Tdom{B_j}}=B_j$ and
    $\Frestr{w_j}{\Tdom{B_j}}=u_i\Comp v_j$.
  \end{itemize}
\end{lemma}
\Beginproof
By induction on the proof $\mu$.

Assume that $\mu$ is an axiom, so that there is a
$k\in\Eset{1,\dots,n}$ such that $A=\Treloc{u_k}{A_k}$, $u_k$ is a
bijection and $\Tdom{A_j}=\emptyset$ for all $j\not=k$. In that case
we have $\Tunderl\mu{\Vect x}=x_k$. There are two subcases to
consider. Assume first that $k=i$. By Lemma~\ref{lemma:reloc} there is
a proof $\rho'$ of
$\Jseq{(\Threloc{B_j}{u_i\Comp v_j})_{j=1}^{n-1}}{\Treloc{u_i}{A_i}}$
such that
$\Tunderl{\rho'}{\Seqext{\Vect x}{i}}=\Tunderl{\rho}{\Seqext{\Vect
    x}{i}}$. We have $C_j=B_{j}$ and $w_j=u_i\Comp v_{j}$ for
$j=1,\dots,n-1$, so that $\rho'$ is a proof of
$\Jseq{\Family{\Threloc{C_j}{w_j}}j1{n-1}}{A}$, so we take $\pi=\rho'$
and equation
$\Tunderl\pi{\Seqext{\Vect x}i}=\Subst{\Tunderl\mu{\Vect
    x}}{\Tunderl\rho{\Seqext{\Vect x}i}}{x_i}$ holds since
$\Tunderl\mu{\Vect x}=x_i$. Assume next that $k\not=i$, then
$\Tdom{A_i}=\emptyset$ and hence $R_j=\Tdom{B_j}=\emptyset$ (and
$v_j=\Fempty\emptyset$) for $j=1,\dots,n-1$. Therefore
$C_j=A_{\Csucc ji}$ and $w_j=v_{\Csucc ji}$ for $j=1,\dots,n-1$. So
our target sequent $\Jseq{\Family{\Threloc{C_j}{w_j}}j1{n-1}}{A}$ can
also be written
$\Jseq{\Family{\Threloc{A_{\Csucc ji}}{u_{\Csucc
        ji}}}j1{n-1}}{\Treloc{u_k}{A_k}}$ and is provable by a proof
$\pi$ such that $\Tunderl{\pi}{\Seqext{\Vect x}i}=x_k$ as contended.

Assume now that $\mu$ is a $\Implies$-intro, that is
$A=\Tfunp{A_{n+1}}{u_{n+1}}{A'}$ and $\mu$ is
\begin{center}
  \AxiomC{$\theta$}
  \noLine
  \UnaryInfC{$\Jseq{\Family{\Threloc{A_j}{u_j}}j1{n+1}}{A'}$}
  \UnaryInfC{$\Jseq{\Family{\Threloc{A_j}{u_j}}j1{n}}{A}$}
  \DisplayProof
\end{center}
We set $B_{n}=\Trestr{A_{n+1}}{\emptyset}$ and of course
$v_{n+1}=\Fempty{\Tdom A}$. Then we have a proof $\rho'$ of
$\Jseq{\Family{\Threloc{B_j}{v_j}}j1n}{A_i}$ such that
$\Tunderl{\rho'}{\Seqext{\Vect
    x}i,x_{n+1}}=\Tunderl{\rho}{\Seqext{\Vect x}i}$ by
Lemma~\ref{lemma:weakening}. We set $C_n=A_{n+1}$ and $w_n=u_n$. Then
by inductive hypothesis applied to $\theta$ (taking $L_n=\Tdom{A_n}$
and $R_n=\emptyset$) we have a proof $\pi^0$ of
$\Jseq{\Family{\Threloc{C_j}{w_j}}j1n}{A'}$ which satisfies
$\Tunderl{\pi^0}{\Seqext{\Vect
    x}i,x_{n+1}}=\Subst{\Tunderl{\theta}{\Vect
    x,x_{n+1}}}{\Tunderl{\rho}{\Seqext{\Vect x}{i}}}{x_i}$ and
applying a $\Implies$-introduction rule we get a proof $\pi$ of
$\Jseq{\Family{\Threloc{C_j}{w_j}}j1{n-1}}{A}$ such that
$\Tunderl\pi{\Seqext{\Vect
    x}i}=\Abs{x_{n+1}}{(\Subst{\Tunderl{\theta}{\Vect
      x,x_{n+1}}}{\Tunderl{\rho}{\Seqext{\Vect
        x}{i}}}{x_i})}=\Subst{\Tunderl\mu{\Vect
    x}}{\Tunderl\rho{\Seqext{\Vect x}i}}{x_i}$ as expected.

Assume last that the proof $\mu$ ends with
\begin{center}
  \AxiomC{$\phi$}
  \noLine
  \UnaryInfC{$\Jseq{\Family{\Threloc{E_j}{s_j}}j1n}{\Tfun{E}{s}{A}}$}
  \AxiomC{$\psi$}
  \noLine
  \UnaryInfC{$\Jseq{\Family{\Threloc{F_j}{t_j}}j1n}{E}$}
  \BinaryInfC{$\Jseq{\Family{\Threloc{A_j}{u_j}}j1n}{A}$}
  \DisplayProof
\end{center}
with $\Tdom{A_j}=\Tdom{E_j}+\Tdom{F_j}$,
$\Trestr{A_j}{\Tdom{E_j}}=E_j$, $\Trestr{A_j}{\Tdom{F_j}}=F_j$,
$\Frestr{u_j}{\Tdom{E_j}}=s_j$ and
$\Frestr{u_j}{\Tdom{F_j}}=s\Comp t_j$, for $j=1,\dots,n$. And we have
$\Tunderl\mu{\Vect x}=\App{\Tunderl\phi{\Vect
    x}}{\Tunderl\psi{\Vect x}}$. The idea is to ``share'' the
substituting proof $\rho$ of
$\Jseq{\Family{\Threloc{B_j}{v_j}}j1n}{A_i}$ among $\phi$ and $\psi$
according to what they need, as specified by the formulas $E_i$ and
$F_i$. So we write $\Tdom{B_j}=L_j+R_j$ where
$L_j=\Funinv{v_j}(\Tdom{E_i})$ and $R_j=\Funinv{v_j}(\Tdom{F_i})$ and
by Lemma~\ref{lemma:restriction} we have two proofs $\rho^L$ of
$\Jseq{\Family{\Threloc{\Trestr{B_j}{L_j}}{v_j^L}}j1{n-1}}{E_i}$ and
$\Jseq{\Family{\Threloc{\Trestr{B_j}{R_j}}{v_j^R}}j1{n-1}}{F_i}$ where we
set $v_j^L=\Frestr{v_j}{L_j}$ and $v_j^R=\Frestr{v_j}{R_j}$, obtained
from $\rho$ by restriction. These proofs satisfy
$\Tunderl{\rho^L}{\Seqext{\Vect x}i}=\Tunderl{\rho^R}{\Seqext{\Vect
    x}i}=\Tunderl{\rho}{\Seqext{\Vect x}i}$.

Now we apply the inductive hypothesis to $\phi$ and $\rho^L$, in order
to get a proof of the sequent
$\Jseq{\Family{\Threloc{G_j}{w_j^L}}j1{n-1}}{\Tfun EsA}$ where
$G_j=\Trestr{C_j}{\Tdom{E_{\Csucc ji}}+L_j}$ (observe indeed that
$\Tdom{E_{\Csucc ji}}\subseteq\Tdom{A_{\Csucc ji}}$ and
$L_j\subseteq\Tdom{B_j}$ and hence are disjoint by our assumption that
$\Tdom{C_j}=\Tdom {A_{\Csucc ji}}+\Tdom{B_j}$) and
$w_j^L=\Frestr{w_j}{\Tdom{E_{\Csucc ji}}+L_j}$. With these
definitions, and by our assumptions about $C_j$ and $w_j$, we have for
all $j=1,\dots,n-1$
\begin{align*}
  \Trestr{G_j}{\Tdom{E_{\Csucc ji}}}
  &=\Trestr{\Trestr{C_j}{\Tdom{A_{\Csucc
    ji}}}}{\Tdom{E_{\Csucc ji}}}=\Trestr{A_{\Csucc
    ji}}{\Tdom{E_{\Csucc ji}}}=E_{\Csucc ji} \\
  \Frestr{w_j^L}{\Tdom{E_{\Csucc ji}}}
  &=\Frestr{\Frestr{w_j}{\Tdom{A_{\Csucc
    ji}}}}{\Tdom{E_{\Csucc ji}}}
    =\Frestr{u_{\Csucc ji}}{\Tdom{E_{\Csucc ji}}}=s_{\Csucc ji}\\
  \Trestr{G_j}{L_j}
  &=\Trestr{\Trestr{C_j}{\Tdom{B_j}}}{L_j}=\Trestr{B_j}{L_j}\\
  \Frestr{w_j^L}{L_j}
  &=\Frestr{\Frestr{w_j}{\Tdom{B_j}}}{L_j}
    =\Frestr{(u_i\Comp v_j)}{L_j}
    =\Frestr{u_i}{\Tdom{E_i}}\Comp v_j^L
    =s_j\Comp v_j^L\,.
\end{align*}
Therefore the inductive hypothesis applies yielding a proof $\phi'$ of
$\Jseq{\Family{\Threloc{G_j}{w_j^L}}j1{n-1}}{\Tfun EsA}$ such that
$\Tunderl{\phi'}{\Seqext{\Vect x}i}=\Subst{\Tunderl\phi{\Vect
    x}}{\Tunderl{\rho^L}{\Seqext{\Vect x}i}}{x_i}=\Subst{\Tunderl\phi{\Vect
    x}}{\Tunderl{\rho}{\Seqext{\Vect x}i}}{x_i}$.

%

Next we apply the inductive hypothesis to $\psi$ and $\rho^R$, in
order to get a proof of the sequent
$\Jseq{\Family{\Threloc{H_j}{r_j}}j1{n-1}}{E}$ where, for $j=1,\dots,n-1$,
$H_j=\Trestr{C_j}{\Tdom{F_{\Csucc ji}}+R_j}$ (again
$\Tdom{F_{\Csucc ji}}\subseteq\Tdom{A_{\Csucc ji}}$ and
$R_j\subseteq\Tdom{B_j}$ are disjoint by our assumption that
$\Tdom{C_j}=\Tdom {A_{\Csucc ji}}+\Tdom{B_j}$) and $r_j$ is defined by
\(
\Frestr{r_j}{\Tdom{F_{\Csucc ji}}}=t_{\Csucc ji}
\) and
\(
  \Frestr{r_j}{R_j}=t_i\Comp v_j^R
\).
Remember indeed that $v_j^R:R_j\to\Tdom{F_i}$ and
$t_i:\Tdom{F_i}\to\Tdom E$. We have
\begin{align*}
  \Trestr{H_j}{\Tdom{F_{\Csucc ji}}}
  &=\Trestr{\Trestr{C_j}{\Tdom{A_{\Csucc
    ji}}}}{\Tdom{F_{\Csucc ji}}}=\Trestr{A_{\Csucc
    ji}}{\Tdom{F_{\Csucc ji}}}=F_{\Csucc ji} \\
  \Trestr{H_j}{R_j}
  &=\Trestr{\Trestr{C_j}{\Tdom{B_j}}}{R_j}=\Trestr{B_j}{R_j}
\end{align*}
and hence by inductive hypothesis there is a proof $\psi'$ of
$\Jseq{\Family{\Threloc{H_j}{r_j}}j1{n-1}}{E}$ such that
$\Tunderl{\psi'}{\Seqext{\Vect x}i}=\Subst{\Tunderl\psi{\Vect
    x}}{\Tunderl{\rho^R}{\Seqext{\Vect x}i}}{x_i}=\Subst{\Tunderl\psi{\Vect
    x}}{\Tunderl{\rho}{\Seqext{\Vect x}i}}{x_i}$.

To end the proof of the lemma, it will be sufficient to prove that we
can apply a $\Implies$-elimination rule to the sequents
$\Jseq{\Family{\Threloc{G_j}{w_j^L}}j1{n-1}}{\Tfun EsA}$ and
$\Jseq{\Family{\Threloc{H_j}{r_j}}j1{n-1}}{E}$ in order to get a proof
$\pi$ of the sequent
$\Jseq{\Family{\Threloc{C_j}{w_j}}j1{n-1}}{A}$. Indeed, the proof
$\pi$ obtained in that way will satisfy
$\Tunderl{\pi}{\Seqext{\Vect x}i}=\App{\Tunderl{\phi'}{\Seqext{\Vect
      x}i}}{\Tunderl{\psi'}{\Seqext{\Vect
      x}i}}=\Subst{\Tunderl\mu{\Vect
    x}}{\Tunderl\rho{\Seqext{\Vect x}i}}{x_i}$. Let
$j\in\Eset{1,\dots,n-1}$. We have $\Trestr{C_j}{\Tdom{G_j}}=G_j$ and
$\Trestr{C_j}{\Tdom{H_j}}=H_j$ simply because $G_j$ and $H_j$ are
defined by restricting $C_j$. Moreover
$\Tdom{G_j}=\Tdom{E_{\Csucc ji}}+L_j$ and
$\Tdom{H_j}=\Tdom{F_{\Csucc ji}}+R_j$. Therefore
$\Tdom{G_j}\cap\Tdom{H_j}=\emptyset$ and
\[
  \Tdom{C_j}=\Tdom{A_{\Csucc ji}}+{\Tdom{B_j}}=\Tdom{E_{\Csucc
      ji}}+\Tdom{F_{\Csucc ji}}+L_j+R_j=\Tdom{G_j}+\Tdom{H_j}\,.
\]
We have $\Frestr{w_j}{\Tdom{G_j}}=w_j^L$ by definition of $w_j^L$ as
$\Frestr{w_j}{\Tdom{E_{\Csucc ji}}+L_j}$. We have
\begin{align*}
  \Frestr{\Frestr{w_j}{\Tdom{H_j}}}{\Tdom{F_{\Csucc ji}}}
  &=\Frestr{\Frestr{w_j}{\Tdom{A_{\Csucc ji}}} }{\Tdom{F_{\Csucc ji}}}
    =\Frestr{u_{\Csucc ji}}{\Tdom{F_{\Csucc ji}}}\\
  &=s\Comp t_{\Csucc ji}
    =\Frestr{(s\Comp r_j)}{\Tdom{F_{\Csucc ji}}}\\
  \Frestr{\Frestr{w_j}{\Tdom{H_j}}}{R_j}
  &=\Frestr{\Frestr{w_j}{\Tdom{B_j}}}{R_j}
    =\Frestr{(u_i\Comp v_j)}{R_j}\\
  &=\Frestr{u_i}{\Tdom{F_i}}\Comp v_j^R
    =s\Comp t_i\Comp v_j^R=s\Comp\Frestr{r_j}{R_j}=\Frestr{(s\Comp r_j)}{R_j}
\end{align*}
and therefore
$\Frestr{w_j}{\Tdom{H_j}}=s\Comp r_j$ as required.
\Endproof

We shall often use the two following consequences of the Substitution Lemma.
\begin{lemma}\label{lemma:substitution-1}
  Given a proof $\mu$ of
  $\Jseq{\Family{\Threloc{A_j}{u_j}}j1n}{A}$ and a proof $\rho$ of
  $\Jseq{\Threloc Bv}{A_i}$ (for some $i\in\Eset{1,\dots,n}$), there
  is a proof $\pi$ of
  $\Jseq{\Family{\Threloc{A_j}{u_j}}j1{i-1}, \Threloc B{u_i\Comp
      v},\Family{\Threloc{A_j}{u_j}}j{i+1}n}{A}$ such that
  $\Tunderl\pi{\Vect x}=\Subst{\Tunderl{\mu}{\Vect x}}{\Tunderl\rho{x_i}}{x_i}$
\end{lemma}
\Beginproof
By weakening we have a proof $\mu'$ of
$\Jseq{\Family{\Threloc{A_j}{u_j}}j1i,
  \Threloc{\Trestr{B}{\emptyset}}{\Fempty{\Tdom A}},
  \Family{\Threloc{A_j}{u_j}}j{i+1}{n}}{A}$ such that
$\Tunderl{\mu'}{\Vect x}=\Tunderl{\mu}{\Seqext{\Vect x}{i+1}}$
(where $\Vect x$ is a list of pairwise distinct variables of length
$n+1$), as well as a proof $\rho'$ of
$\Jseq{\Family{\Threloc{\Trestr{A_j}\emptyset}{\Fempty{\Tdom{A_i}}}}j1i,
  \Threloc{B}{v},
  \Family{\Threloc{\Trestr{A_j}\emptyset}{\Fempty{\Tdom{A_i}}}}j{i+1}{n}}{A_i}$
such that $\Tunderl{\rho'}{\Vect x}=\Tunderl\rho{x_{i+1}}$. By
Lemma~\ref{lemma:substitution}, we have a proof $\pi'$ of
$\Jseq{\Family{\Threloc{A_j}{u_j}}j1{i-1}, \Threloc B{u_i\Comp
    v},\Family{\Threloc{A_j}{u_j}}j{i+1}n}{A}$ which satisfies
$\Tunderl{\pi'}{\Seqext{\Vect x}i}=\Subst{\Tunderl{\mu'}{\Vect
    x}}{\Tunderl{\rho'}{\Seqext{\Vect
      x}{i}}}{x_i}=\Subst{\Tunderl{\mu}{\Vect
    x}}{\Tunderl\rho{x_i}}{x_i}$.
\Endproof

\begin{lemma}\label{lemma:substitution-2}
  Given a proof $\mu$ of $\Jseq{\Threloc Av}{B}$ and a proof
  $\rho$ of $\Jseq{\Family{\Threloc{A_j}{u_j}}j1n}{A}$, there is a
  proof $\pi$ of $\Jseq{\Family{\Threloc{A_j}{v\Comp u_j}}j1{n}}{B}$
  such that
  $\Tunderl\pi{\Vect
    x}=\Subst{\Tunderl{\mu}{x}}{\Tunderl\rho{\Vect x}}{x}$.
\end{lemma}
The proof is similar to the previous one.

If $A$ and $B$ are formulas such that $\Tunder A=\Tunder B$,
$\Tdom A=\Tdom B$ and $\Tfam A=\Tfam B$, we say that $A$ and $B$ are
similar and we write $A\Tsim B$. One fundamental property of our
deduction system is that two formulas which represent the same family
of intersection types are logically equivalent.
\begin{theorem}\label{th:Tsim-implies}
  If $A\Tsim B$ then $\Jseq{\Threloc A\Id}{B}$ with a proof $\pi$ such
  that $\Tunderl\pi x\Etaeq x$.
\end{theorem}
\Beginproof
Assume that $A=\Tcst\alpha f$, then we have $B=A$ and
$\Jseq{\Threloc A\Id}{B}$ is an axiom.

Assume that $A=\Tfunp CuD$ and $B=\Tfunp EvF$.  We have $D\Tsim F$ and
hence $\Jseq{\Threloc{D}{\Id}}{F}$ with a proof $\rho$ such that
$\Tunderl\rho x\Etaeq x$. And there is a bijection
$w:\Tdom E\to\Tdom C$ such that $\Treloc wE\Tsim C$ and $u\Comp
w=v$. By inductive hypothesis we have a proof $\mu$ of
$\Jseq{\Threloc{\Treloc wE}\Id}{C}$ such that
$\Tunderl\mu y\Etaeq y$, and hence using the axiom
$\Jseq{\Threloc Ew}{\Treloc wE}$ and Lemma~\ref{lemma:substitution-1}
we have a proof $\mu'$ of $\Jseq{\Threloc Ew}{C}$ such that
$\Tunderl{\mu'}x=\Tunderl{\mu}x$.

There is a proof $\pi^1$ of
$\Jseq{\Threloc{\Tfunp{C}{u}{D}}\Id,\Threloc Cu}{D}$ such that
$\Tunderl{\pi^1}{x,y}=\App xy$ (consider the two axioms
$\Jseq{\Threloc{\Tfunp{C}{u}{D}}\Id, \Threloc{\Trestr
    C\emptyset}{\Fempty{\Tdom D}}}{\Tfun CuD}$ and
$\Jseq{\Threloc{\Trestr{\Tfunp{C}{u}{D}}{\emptyset}}{\Fempty{\Tdom
      C}}, \Threloc C\Id}{C}$ and use a $\Implies$-elimination rule). So by
Lemma~\ref{lemma:substitution-1} there is a proof $\pi^2$ of
$\Jseq{\Threloc{\Tfunp{C}{u}{D}}\Id,\Threloc E{u\Comp w}}{D}$, that is
of $\Jseq{\Threloc{\Tfunp{C}{u}{D}}\Id,\Threloc E{v}}{D}$, such that
$\Tunderl{\pi^2}{x,y}=\App x{\Tunderl\mu y}$. Applying
Lemma~\ref{lemma:substitution-2} we get a proof $\pi^3$ of
$\Jseq{\Threloc{\Tfunp CuD}\Id,\Threloc{E}{v}}{F}$ such that
$\Tunderl{\pi^3}{x,y}=\Subst{\Tunderl\rho z}{\App x{\Tunderl\mu y}}z$.
We get the expected proof $\pi$ by a $\Implies$-introduction rule so that
$\Tunderl\pi x=\Abs y{\Subst{\Tunderl\rho z}{\App x{\Tunderl\mu y}}z}$.
By inductive hypothesis $\Tunderl\pi x \Etaeq x$.
\Endproof

\subsection{Relation between intersection types and $\LJ(I)$}

Now we explain the precise connection between non-idempotent
intersection types and our logical system $\LJ(I)$. This connection
consists of two statements:
\begin{itemize}
\item the first one means that any proof of $\LJ(I)$ can be seen as a
  typing derivation in non-idempotent intersection types (soundness)
\item and the second one means that any non-idempotent intersection
  typing can be seen as a derivation in $\LJ(I)$ (completeness).
\end{itemize}

\begin{theorem}[Soundness]\label{th:simple-soundness}
  Let $\pi$ be a deduction tree of the sequent
  $\Jseq{\Family{\Threloc {A_i}{u_i}}i1n}{B}$ and
  $\Vect x$ a sequence of $n$ pairwise distinct variables. Then the
  $\lambda$-term $\Tunderl\pi{\Vect x}$ satisfies
  \(
    \Tseq{\Family{x_i:\Tfam{\Threloc{A_i}{u_i}}_j:\Tunder{A_i}}i1n}
    {\Tunderl\pi{\Vect x}}{\Tfam B_j:\Tunder B}
  \)
  in the intersection type system, for each $j\in\Tdom B$.
\end{theorem}
\Beginproof
We prove the first part by induction on $\pi$ (in the course of this
induction, we recall the precise definition of $\Tunderl\pi{\Vect x}$). If
$\pi$ is the proof
\begin{center}
  \AxiomC{$q\not=i\Implies\Tdom{A_q}=\emptyset$ and $u_i$ is a bijection}
  \UnaryInfC{$\Jseq{\Family{\Threloc {A_q}{u_q}}q1n}
    {\Treloc{u_i}{A_i}}$}
  \DisplayProof
\end{center}
(so that $B=\Treloc{u_i}{A_i}$) then $\Tunderl\pi{\Vect x}=x_i$.  We
have $\Tfam{\Threloc{A_q}{u_q}}_j=\Mset{}$ if $q\not=i$,
$\Tfam{\Threloc{A_i}{u_i}}_j=\Mset{\Tfam{A_i}_{\Funinv{u_i}(j)}}$ and
$\Tfam{\Treloc{u_i}{A_i}}_j=\Tfam{A_i}_{\Funinv{u_i}(j)}$. It follows that
$\Tseq{\Family{x_q:\Tfam{\Threloc{A_q}{u_q}}_j:\Tunder{A_q}}q1n}
    {x_i}{\Tfam{B}_j:\Tunder{B}}$
is a valid axiom in the intersection type system.

Assume that $\pi$ is the proof
\begin{center}
  \noLine
  \AxiomC{$\pi^0$}
  \UnaryInfC{$\Jseq{\Threloc {A_1}{u_1},\dots,\Threloc {A_n}{u_n},\Threloc Au}
    {B}$}
  \UnaryInfC{$\Jseq{\Threloc {A_1}{u_1},\dots,\Threloc {A_n}{u_n}}
    {\Tfun{A}{u}B}$}
  \DisplayProof
\end{center}
where $\pi^0$ is the proof of the premise of the last rule of
$\pi$. By inductive hypothesis the $\lambda$-term
$\Tunderl{\pi^0}{\Vect x,x}$ satisfies
\(
  \Tseq{\Family{x_i:\Tfam{\Threloc{A_i}{u_i}}_j:\Tunder{A_i}}i1n,
  x:\Tfam{\Threloc Au}_j:\Tunder A}
  {\Tunderl{\pi^0}{\Vect x,x}}{\Tfam B_j:\Tunder B}
\)
from which we deduce
\(
  \Tseq{\Family{x_i:\Tfam{\Threloc{A_i}{u_i}}_j:\Tunder{A_i}}i1n}
  {\Abst{x}{\Tunder{A}}{\Tunderl{\pi^0}{\Vect x,x}}}
  {(\Tfam{\Threloc{A}{u}}_j,\Tfam B_j):
  \Timpl{\Tunder A}{\Tunder B}}  
\)
which is the required judgment since
$\Tunderl\pi{\Vect x}=\Abst{x}{\Tunder{A}}{\Tunderl{\pi^0}{\Vect
    x,x}}$ and
$(\Tfam{\Threloc{A_i}{u_i}}_j,\Tfam B_j)=\Tfam{\Tfun{A}{u}{B}}_j$
as easily checked.

Assume last that $\pi$ ends with
\begin{center}
  \noLine
  \AxiomC{$\pi^1$}
  \UnaryInfC{$\Jseq{\Threloc {C_1}{u_1},\dots,\Threloc {C_n}{u_n}}
    {\Tfun AuB}$}
  \noLine
  \AxiomC{$\pi^2$}
  \UnaryInfC{$\Jseq{\Threloc {D_1}{v_1},\dots,\Threloc {D_n}{v_n}}
    {A}$}
  \BinaryInfC{$\Jseq{\Threloc {E_1}{w_1},\dots,\Threloc {E_n}{w_n}}
    {B}$}
  \DisplayProof
\end{center}
with: for each $i=1,\dots,n$ there are two disjoint sets
$L_i$ and $R_i$ such that $\Tdom{E_i}=L_i+ R_i$,
$C_i=\Trestr{E_i}{L_i}$, $D_i=\Trestr{E_i}{R_i}$,
$\Funrestr{w_i}{L_i}=u_i$, and $\Funrestr{w_i}{R_i}=u\Comp v_i$.

Let $j\in\Tdom B$.  By inductive hypothesis, the judgment
$\Tseq{\Family{x_i:\Tfam{\Threloc{C_i}{u_i}}_j:\Tunder{C_i}}i1n}
{\Tunderl{\pi^1}{\Vect x}}{\Tfam{\Tfun{A}{u}{B}}_j}$ is derivable in
the intersection type system.  Let $K_j=\Funinv u(\Eset j)$, which is
a finite subset of $\Tdom A$. By inductive hypothesis again, for each
$k\in K_j$ we have
\(
  \Tseq{\Family{x_i:\Tfam{\Threloc{D_i}{u_i}}_k:\Tunder{D_i}}i1n}
  {\Tunderl{\pi^2}{\Vect x}}{\Tfam{A}_k}\,.
\)
Now observe that
$\Tfam{\Tfun AuB}_j=(\Mset{\Tfam A_k\St k\in K_j},\Tfam B_j)$ so that
\begin{align*}
  \Tseq{\Family{x_i:\Tfam{\Threloc{C_i}{u_i}}_j+\sum_{k\in
  K_j}\Tfam{\Threloc{D_i}{u_i}}_k:\Tunder{E_i}}i1n}{\App{\Tunderl{\pi^1}{\Vect x}}{\Tunderl{\pi^2}{\Vect x}}}
  {\Tfam B_j}
\end{align*}
is derivable in intersection types (remember that
$\Tunder{C_i}=\Tunder{D_i}=\Tunder{E_i}$). Since
$\Tunderl\pi{\Vect x}=\App{\Tunderl{\pi^1}{\Vect
    x}}{\Tunderl{\pi^2}{\Vect x}}$ it will be sufficient to prove that
\begin{align}\label{eq:app-formulas-multisets}
  \Tfam{\Threloc{E_i}{w_i}}_j=\Tfam{\Threloc{C_i}{u_i}}_j+\sum_{k\in
  K_j}\Tfam{\Threloc{D_i}{v_i}}_k\,.
\end{align}
For this, since
$\Tfam{\Threloc{E_i}{w_i}}_j=\Mset{\Tfam{E_i}_l\St w_i(l)=j}$,
consider an element $l$ of $\Tdom{E_i}$ such that $w_i(l)=j$. There
are two possibilities:
(1)~either $l\in L_i$ and in that case we know that
  $\Tfam{E_i}_l=\Tfam{C_i}_l$ since $\Trestr{E_i}{L_i}=C_i$ and
  moreover we have $u_i(l)=w_i(l)=j$
  (2)~or $l\in R_i$. In that case we have $\Tfam{E_i}_l=\Tfam{D_i}_l$
  since $\Trestr{E_i}{R_i}=D_i$. Moreover $u(v_i(l))=w_i(l)=j$  and
  hence $v_i(l)\in K_j$.
Therefore
\begin{align*}
  \Mset{\Tfam{E_i}_l\St l\in L_i\text{ and }w_i(l)=j}
    &=\Mset{\Tfam{C_i}_l\St u_i(l)=j}=\Tfam{\Threloc{C_i}{u_i}}_j\\
  \Mset{\Tfam{E_i}_l\St l\in R_i\text{ and }w_i(l)=j}
    &=\Mset{\Tfam{D_i}_l\St v_i(l)\in K_j} =\sum_{k\in
      K_j}\Tfam{\Threloc{D_i}{v_i}}_k
\end{align*}
and~\Eqref{eq:app-formulas-multisets} follows.
\Endproof

\begin{theorem}[Completeness]\label{th:simple-completeness}
  Let $J\subseteq I$. Let $M$ be a $\lambda$-term and $\List x1n$ be
  pairwise distinct variables, such that
  \(
    \Tseq{\Family{x_i:m^j_i:\sigma_i}i1n}{M}{b_j:\tau}
  \)
  in the intersection type system for all $j\in J$.
  Let $\List A1n$ and $B$ be formulas and let $\List u1n$ be almost
  injective functions such that $u_i:\Tdom{A_i}\to J=\Tdom B$. Assume
  also that $\Tunder{A_i}=\sigma_i$ for each $i=1,\dots,n$ and that
  $\Tunder B=\tau$. Last assume that, for all $j\in J$, one has
  $\Tfam B_j=b_j$ and $\Tfam{\Threloc{A_i}{u_i}}_j=m_i^j$ for
  $i=1,\dots,n$. Then the judgment
  \(
    \Jseq{\Family{\Threloc{A_i}{u_i}}i1n}{B}
  \)
  has a proof $\pi$ such that $\Tunderl\pi{\Vect x}\Etaeq M$.
\end{theorem}
\Beginproof
By induction on $M$. Assume first that $M=x_i$ for some
$i\in\Eset{1,\dots,n}$. Then we must have $\tau=\sigma_i$,
$m_q^j=\Mset{}$ for $q\not=i$ and $m_i^j=\Mset{b_j}$ for all $j\in
J$. Therefore $\Tdom{A_q}=\emptyset$ and $u_q$ is the empty function
for $q\not=i$, $u_i$ is a bijection $\Tdom{A_i}\to J$ and
$\forall k\in\Tdom{A_i}\ \Tfam{A_i}_k=b_{u_i(k)}$, in other words
$\Treloc{u_i}{A_i}\Tsim B$. By Theorem~\ref{th:Tsim-implies} we know
that the judgment $\Jseq{\Threloc{(\Treloc{u_i}{A_i})}{\Id}}{B}$ is
provable in $\LJ(I)$ with a proof $\rho$ such that
$\Tunderl\rho x\Etaeq x$. We have a proof $\theta$ of
$\Jseq{\Family{\Threloc{A_i}{u_i}}i1n}{\Treloc{u_i}{A_i}}$ which
consists of an axiom so that $\Tunderl\theta{\Vect x}=x_i$ and hence
by Lemma~\ref{lemma:substitution-2} we have a proof $\pi$ of
$\Jseq{\Family{\Threloc{A_i}{u_i}}i1n}{B}$ such that
$\Tunderl\pi{\Vect x}=\Subst{\Tunderl\rho x}{\Tunderl\theta{\Vect
    x}}x\Etaeq x_i$.

Assume that $M=\Abst{x}{\sigma}{N}$, that $\tau=(\Timpl\sigma\phi)$ and
that we have a family of deductions (for $j\in J$) of
\(
  \Tseq{\Family{x_i:m^j_i:\sigma_i}i1n}{M}
  {(m^j,c_j):\Timpl\sigma\phi}
\)
with $b_j=(m^j,c_j)$ and the premise of this conclusion in each of
these deductions is
\(
  \Tseq{\Family{x_i:m^j_i:\sigma_i}i1n,x:m^j:\sigma}{N}
  {c_j:\phi}
\).
We must have $B=\Tfunp CuD$ with $\Tunder D=\phi$, $\Tunder C=\sigma$,
$\Tdom D=J$, $u:\Tdom C\to\Tdom D$ almost injective, $\Tfam D_j=c_j$
and $\Mset{\Tfam C_k\St k\in\Tdom C\text{ and }u(k)=j}=m^j$, that is
$\Tfam{\Threloc Cu}_j=m^j$, for each $j\in J$. By inductive hypothesis
we have a proof $\rho$ of
\(
  \Jseq{\Family{\Threloc{A_i}{u_i}}i1n,\Threloc Cu}{D}
\)
such that $\Tunderl\rho{\Vect x,x}\Etaeq N$ from which we obtain a
proof $\pi$ of
$\Jseq{\Family{\Threloc{A_i}{u_i}}i1n}{\Tfun CuD}$ such
that
$\Tunderl\pi{\Vect x}=\Abst x\sigma{\Tunderl\rho{\Vect x,x}}\Etaeq
M$ as expected.

Assume last that $M=\App NP$ and that we have a $J$-indexed family of
deductions
\(
  \Tseq{\Family{x_i:m^j_i:\sigma_i}i1n}{M}{b_j:\tau}\,.  
\)
Let $\List A1n$, $\List u1n$ and $B$ be $\LJ(I)$ formulas and almost
injective functions as in the statement of the theorem.

Let $j\in J$. There is a finite set $L_j\subseteq I$ and multisets
$m_i^{j,0}$, $(m_i^{j,l})_{l\in L_j}$ such that we have
deductions\footnote{Notice that our $\lambda$-calculus is in
  \emph{Church style} and hence the type $\sigma$ is uniquely
  determined by the sub-term $N$ of $M$.} of
\(
  \Tseq{\Family{x_i:m^{j,0}_i:\sigma_i}i1n}{N}
  {(\Mset{a^j_l\St l\in L_j},b_j):\Timpl\sigma\tau}
\)
and, for each $l\in L_j$, of
\(
  \Tseq{\Family{x_i:m^{j,l}_i:\sigma_i}i1n}{P}{a^j_l:\sigma}
\)
with
\begin{align}\label{eq:mset-hyp-decomp}
  m_i^j=m_i^{j,0}+\sum_{l\in L_j}m_i^{j,l}\,.
\end{align}
We assume the finite sets $L_j$ to be pairwise disjoint (this is
possible because $I$ is infinite) and we use $L$ for their union. Let
$u:L\to J$ be the function which maps $l\in L$ to the unique $j$ such
that $l\in L_j$, this function is almost injective. Let $A$ be an
$\LL(J)$ formula such that $\Tunder A=\sigma$, $\Tdom A=L$ and
$\Tfam A_l=a_l^{u(l)}$; such a formula exists by
Proposition~\ref{prop:family-rep}.

Let $i\in\Eset{1,\dots,n}$. For each $j\in J$ we know that
\[
  \Mset{\Tfam{A_i}_r\St r\in\Tdom{A_i}\text{ and }u_i(r)=j}=m_i^j
  =m_i^{j,0}+\sum_{l\in L_j}m_i^{j,l}
\]
and hence we can split the set $\Tdom{A_i}\cap\Funinv{u_i}(\Eset{j})$
into disjoint subsets $R_i^{j,0}$ and $(R_i^{j,l})_{l\in L_j}$ in such
a way that
\[
  \Mset{\Tfam{A_i}_r\St r\in R_i^{j,0}}=m_i^{j,0}
  \text{\quad and\quad}
  \forall l\in L_j\ \Mset{\Tfam{A_i}_r\St r\in R_i^{j,l}}=m_i^{j,l}\,.
\]
We set $R_i^0=\Union_{j\in J}R_i^{j,0}$; observe that this is a
disjoint union because
$R_i^{j,0}\subseteq\Funinv{u_i}(\Eset{j})$. Similarly we define
$R_i^1=\Union_{l\in L}R_i^{u(l),l}$ which is a disjoint union for the
following reason: if $l,l'\in L$ satisfy $u(l)=u(l')=j$ then
$R_i^{j,l}$ and $R_i^{j,l'}$ have been chosen disjoint and if $u(l)=j$
and $u(l')=j'$ with $j\not=j'$ we have
$R_i^{j,l}\subseteq\Funinv{u_i}{\Eset j}$ and
$R_i^{j',l'}\subseteq\Funinv{u_i}{(\Eset{j'})}$.  Let $v_i:R_i^1\to L$
be defined by: $v_i(r)$ is the unique $l\in L$ such that
$r\in R_i^{u(l),l}$. Since each $R_i^{j,l}$ is finite the function
$v_i$ is almost injective. Moreover $u\Comp
v_i=\Frestr{u_i}{R_i^1}$.

We use $u'_i$ for the restriction of $u_i$ to $R_i^0$ so that
$u'_i:R_i^0\to J$.  By inductive hypothesis we have
\(
  \Jseq{\Family{\Threlocp{\Trestr{A_i}{R_i^0}}{u'_i}}i1n}{\Tfun{A}{u}{B}}
\)
with a proof $\mu$ such that
$\Tunderl\mu{\Vect x}\Etaeq N$. Indeed
$\Mset{\Tfam{\Trestr{A_i}{R_i^0}}_r\St r\in R_i^0\text{ and
  }u'_i(r)=j}=m_i^{j,0}$ and
$\Tfam{\Tfun AuB}_j=(\Mset{a_l^j\St u(l)=j},b_j)$ for each $j\in J$.
For the same reason we have
\(
  \Jseq{\Family{\Threlocp{\Trestr{A_i}{R_i^1}}{v_i}}i1n}{A}
\)
with a proof $\rho$ such that $\Tunderl\rho{\Vect x}\Etaeq
P$. Indeed for each $l\in L=\Tdom A$ we have
\(
  \Mset{\Tfam{\Trestr{A_i}{R_i^1}}_r\St v_i(r)=l}=m_i^{j,l}
\)
and $\Tfam A_l=a_l^{j}$ where $j=u(l)$. By an application rule we get
a proof $\pi$ of
\(
  \Jseq{\Family{\Threloc{A_i}{u_i}}i1n}{B}
\)
such that
$\Tunderl\pi{\Vect x}=\App{\Tunderl\mu{\Vect
    x}}{\Tunderl\rho{\Vect x}}\Etaeq\App NP=M$ as contended.
\Endproof


\section{The untyped Scott case}\label{sec:D-infty-relational}
Since intesection types usually apply to the pure $\lambda$-calculus,
we move now to this setting by choosing in $\RELK$ the set $\Dinfr$
as model of the pure $\lambda$-calculus.  The $\Dinfr$ intersection
typing system has the elements of $\Dinfr$ as types, and the typing
rules involve sequents of shape $\Tseq{\Family{x_i:m_i}i1n}Ma$ where
$m_i\in\Mset{\Dinfr}$ and $a\in\Dinfr$.

We use $\Lamt$ for the set of terms of the pure $\lambda$-calculus,
and $\Lamto$ as the pure $\lambda$-calculus extended with a constant
$\Tinflam$ subject to the two following $\Inflam$ reduction rules:
$\Abs x\Tinflam \Rel\Inflam \Tinflam$ and
$\App\Tinflam M \Rel\Inflam \Tinflam$.  We use $\Etaoeq$ for the least
congruence on $\Lamto$ which contains $\Eta$ and $\Inflam$ and
similarly for $\Beoeq$.  We define a family $(\Qproj x)_{x\in\Varset}$
of subsets of $\Lamto$ minimal such that, for any sequence
$\Vect x=(\List x1n)$ and $\Vect y=(\List y1k)$ such that
$\Vect x,\Vect y$ is repetition-free, and for any terms
$M_i\in\Qproj{x_i}$ (for $i=1,\dots,n$), one has
\( \Abs{\Vect x}{\Abs{\Vect y}{\App x{M_1\cdots
      M_n\Appsep\Tinflam\cdots\Tinflam}}}\in\Qproj x \).  Notice that
all the elements of $\Qproj x$ are normal and that $x\in\Qproj x$.


The typing rules are
\begin{center}
  \AxiomC{}
  \UnaryInfC{$\Tseq{x_1:\Mset{},\dots,x_i:\Mset a,\dots,x_n:\Mset{}}{x_i}{a}$}
  \DisplayProof
\quad
  \AxiomC{$\Tseq{\Phi,x:m}{M}{a}$}
  \UnaryInfC{$\Tseq{\Phi}{\Abs xM}{(m,a)}$}
  \DisplayProof
\end{center}
\begin{center}
  \AxiomC{$\Tseq{\Phi}{M}{(\Mset{\List a1k},b)}$}
  \AxiomC{$(\Tseq{\Phi_j}{N}{a_j})_{j=1}^k$}
  \BinaryInfC{$\Tseq{\Phi+\sum_{j=1}^k\Phi_j}{\App MN}{b}$}
  \DisplayProof
\end{center}
where we use the following convention: when we write $\Phi+\Psi$ it is
assume that $\Phi$ is of shape $\Family{x_i:m_i}i1n$ and $\Psi$
is of shape $\Family{x_i:p_i}i1n$, and then $\Phi+\Psi$ is
$\Family{x_i:m_i+p_i}i1n$.
%
%
This typing system is just a ``proof-theoretic'' rephrasing of the
denotational semantics of the terms of $\Lamto$ in $\Dinfr$.
\begin{proposition}
  Let $M,M'\in\Lamto$ and $\Vect x=(\List x1n)$ be a list of pairwise
  distinct variables containing all the free variables of $M$ and
  $M'$. Let $m_i\in\Mfin{\Dinfr}$ for $i=1,\dots,n$ and $b\in\Dinfr$.
  If $M\Beoeq M'$ then $\Tseq{\Family{x_i:m_i}i1n}{M}{b}$
  iff $\Tseq{\Family{x_i:m_i}i1n}{M'}{b}$.
\end{proposition}

\subsection{Formulas}
We define the associated formulas as follows, each formula $A$ being
given together with $\Tdom A\subseteq I$ and $\Tfam A\in\Dinfr^J$.
\begin{itemize}
\item If $J\subseteq I$ then $\Estackf J$ is a formula with
  $\Tdom{\Estackf J}=J$ and $\Tfam{\Estackf J}_j=\Estack$ for $j\in J$
\item and if $A$ and $B$ are formulas and $u:\Tdom A\to\Tdom B$ is
  almost injective then $\Tfun AuB$ is a formula with
  $\Tdom{\Tfun AuB}=\Tdom B$ and
  $\Tfam{\Tfun AuB}_j=(\Mset{\Tfam A_k\St u(k)=j},\Tfam
  B_j)\in\Dinfr$.
\end{itemize}
We can consider that there is a type $\Otype$ of pure $\lambda$-terms
interpreted as $\Dinfr$ in $\RELK$, such that
$\Timplp\Otype\Otype=\Otype$, and then for any formula $A$ we have
$\Tunder A=\Otype$.

Operations of restriction and relocation of formulas are the same as
in Section~\ref{sec:simple-types} (setting
$\Trestr{\Estackf J}{K}=\Estackf{J\cap K}$) and satisfy the same
properties, for instance $\Tfam{\Trestr AK}=\Frestr{\Tfam A}K$ and one
sets $\Treloc u{\Estackf J}=\Estackf K$ if $u:J\to K$ is a bijection.

The deduction rules are exactly the same as those of
Section~\ref{sec:simple-types}, plus the following axiom:
\begin{center}
  \AxiomC{}
  \UnaryInfC{$\Jseq{}{\Estackf\emptyset}$}
  \DisplayProof
\end{center}
With any deduction $\pi$ of $\Jseq{\Family{\Threloc{A_i}{u_i}}i1n}{B}$
and sequence $\Vect x=(\List x1n)$ of pairwise distinct variables, we
can associate a \emph{pure} $\Tunderl\pi{\Vect x}\in\Lamto$ defined
exactly as in Section~\ref{sec:simple-types} (just drop the types
associated with variables in abstractions). If $\pi$ consists of an
instance of the additional axiom, we set
$\Tunderl\pi{\Vect x}=\Tinflam$.

\begin{lemma}\label{lemme:empty-proofs}
  Let $A,\List A1n$ be a formula such that
  $\Tdom A=\Tdom{A_i}=\emptyset$. Then
  $\Jseq{\Family{\Threloc{A_i}{\Fempty\emptyset}}i1n}{A}$ is provable
  by a proof $\pi$ which satisfies
  $\Tunderl\pi{\List x1k}\Omegaeq\Tinflam$.
\end{lemma}
The proof is a straightforward induction on $A$ using the additional
axiom, Lemma~\ref{lemma:weakening} and the observations that if
$\Tdom{\Tfun BuC}=\emptyset$ then $u=\Fempty\emptyset$.

One can easily define a size function $\Dinfrsize:\Dinfr\to\Nat$ such
that $\Dinfrsize(\Estack)=0$ and
$\Dinfrsize(\Mset{\List
  a1k},a)=\Dinfrsize(a)+\sum_{i=1}^k(1+\Dinfrsize(a_i))$.
First we have to prove an adapted version of
Proposition~\ref{prop:family-rep}; here it will be restricted to finite
sets.

%

\begin{proposition}\label{prop:dinf-family-rep}
  Let $J$ be a \emph{finite} subset of $I$ and $f\in\Dinfr^J$. There
  is a formula $A$ such that $\Tdom A=J$ and $\Tfam A=f$.
\end{proposition}
\Beginproof
Observe that, since $J$ is finite, there is an $N\in\Nat$ such that
$\forall j\in J\ \forall q\in\Nat\ q\geq N\Implies f(j)_q=\Mset{}$
(remember that $f(j)\in\Mfin{\Dinfr}^\Nat$). Let $N(f)$ be the least
such $N$.  We set $\Dinfrsize(f)=\sum_{j\in J}\Dinfrsize(f(j))$ and
the proof is by induction on $(\Dinfrsize(f),N(f))$ lexicographically.

If $\Dinfrsize(f)=0$ this means that $f(j)=\Estack$ for all $j\in J$
and hence we can take $A=\Estackf J$. Assume that $\Dinfrsize(f)>0$,
one can write\footnote{This is also possible if $\Dinfrsize(f)=0$
  actually.}  $f(j)=(m_j,a_j)$ with $m_j\in\Mfin{\Dinfr}$ and
$a_j\in\Dinfr$ for each $j\in J$. Just as in the proof of
Proposition~\ref{prop:family-rep} we choose a set $K$, a function
$g:K\to\Dinfr$ and an almost injective function $u:K\to J$ such that
$m_j=\Mset{g(k)\St u(k)=j}$. The set $K$ is finite since $J$ is and we
have $\Dinfrsize(g)<\Dinfrsize(f)$ because
$\Dinfrsize(f)>0$. Therefore by inductive hypothesis there is a 
formula $B$ such that $\Tdom B=K$ and $\Tfam B=g$. Let $f':J\to\Dinfr$
defined by $f'(j)=a_j$, we have $\Dinfrsize(f')\leq\Dinfrsize(f)$ and
$N(f')<N(f)$ and hence by inductive hypothesis there is a 
formula $C$ such that $\Tfam C=g$. We set $A=\Tfunp BuC$ which
satisfies $\Tfam A=f$ as required. 
\Endproof

Theorem~\ref{th:Tsim-implies} still holds up to some mild
adaptation. First notice that $A\Tsim B$ simply means now that
$\Tdom A=\Tdom B$ and $\Tfam A=\Tfam B$.




\begin{theorem}\label{th:Dinfr-Tsim-implies}
  If $A$ and $B$ are
  such that $A\Tsim B$ then $\Jseq{\Threloc A\Id}{B}$ with a proof
  $\pi$ which satisfies $\Tunderl\pi x\in\Qproj x$.
\end{theorem}
\Beginproof
By induction on the sum of the sizes of $A$ and $B$.  Assume that
$A=\Estackf J$ so that $\Tdom B=J$ and
$\forall j\in J\ \Tfam B_j=\Estack$. There are two cases as to $B$. In
the first case $B$ is of shape $\Estackf K$ but then we must have
$K=J$ and we can take for $\pi$ and axiom so that
$\Tunderl\pi x=x\in\Qproj x$. Otherwise we have $B=\Tfunp CuD$ with
$\Tdom D=J$, $\forall j\in J\ \Tfam D_j=\Estack$ and
$\Tdom C=\emptyset$, so that $u=\Fempty J$. We have $A\Tsim D$ and
hence we have a proof $\rho$ of $\Jseq{\Threloc A\Id}{D}$ such that
$\Tunderl\rho x\in\Qproj x$. By weakening and $\Implies$-introduction
we get a proof $\pi$ of $\Jseq{\Threloc A\Id}{B}$ which satisfies
$\Tunderl\pi x=\Abs y{\Tunderl\rho x}$. Notice that in this case the
proof $\rho$ will be of the same shape as $\rho$ so that we will have
$\Tunderl\pi{x}=\Abs{\Vect y}x\in\Qproj x$.

Assume that $A=\Tfunp CuD$. If $B=\Estackf J$ then we must have
$\Tdom C=\emptyset$, $u=\Fempty J$ and $D\Tsim B$ and hence by
inductive hypothesis we have a proof $\rho$ of
$\Jseq{\Threloc D\Id}{B}$ such that $\Tunderl\rho x\in\Qproj x$, from
which using the axiom $\Jseq{}{\Estackf\emptyset}$, we build a proof
$\pi$ of $\Jseq{\Threloc A\Id}{B}$ such that
$\Tunderl\pi x=\Subst{\Tunderl\rho y}{\App x\Tinflam}{y}\in\Qproj x$.
Notice that in this case the proof $\rho$ will be of the same shape as
$\rho$ so that we will have
$\Tunderl\pi{x}=\App x{\Vect \Tinflam}\in\Qproj x$ (where
$\Vect\Tinflam$ is a finite sequence of terms $\Tinflam$).

Assume last that $B=\Tfunp EvF$, then we must have $D\Tsim F$ and
there must be a bijection $w:\Tdom E\to\Tdom C$ such that $u\Comp w=v$
and $\Treloc wE\Tsim C$. We reason as in the proof of
Lemma~\ref{th:Tsim-implies}: by inductive hypothesis we have a proof
$\rho$ of $\Jseq{\Threloc D\Id}{F}$ and a proof $\mu$ of
$\Jseq{\Threloc{\Treloc wE}\Id}{C}$ from which we build a proof $\pi$
of $\Jseq{\Threloc A\Id}{B}$ such that
$\Tunderl\pi x=\Abs y{\Subst{\Tunderl\rho z}{\App x{\Tunderl\mu
      y}}{z}}\in\Qproj x$ by inductive hypothesis.
\Endproof

\begin{theorem}[Soundness]\label{th:Dinfr-soundness}
  Let $\pi$ be a deduction tree of the sequent
  $\Jseq{\Threloc {A_1}{u_1},\dots,\Threloc {A_n}{u_n}} {B}$ and
  $\Vect x$ a sequence of $n$ pairwise distinct variables. Then the
  $\lambda$-term $\Tunderl\pi{\Vect x}\in\Lamto$ satisfies
  \(
    \Tseq{\Family{x_i:\Tfam{\Threloc{A_i}{u_i}}_j}i1n}
    {\Tunderl\pi{\Vect x}}{\Tfam B_j}
  \)
  in the $\Dinfr$ intersection type system, for each $j\in\Tdom B$.
\end{theorem}
The proof is exactly the same as that of
Theorem~\ref{th:simple-soundness}, dropping all simple types.

For all $\lambda$-term $M\in\Lamt$, we define $\Qprojo M$ as the least subset of
element of $\Lamto$ such that:
\begin{itemize}
\item if $O\in\Lamto$ and $O\Omegaeq\Tinflam$ then $O\in\Qprojo M$ for
  all $M\in\Lamt$
\item if $M=x$ then $\Qproj x\subseteq\Qprojo M$
\item if $M\in\Abs yN$ and $N'\in\Qprojo N$ then $\Abs y{N'}\in\Qprojo M$
\item if $M=\App NP$, $N'\in\Qprojo N$ and $P'\in\Qprojo P$ then
  $\App{N'}{P'}\in\Qprojo M$.
\end{itemize}
The elements of $\Qprojo M$ can probably be seen as approximates of
$M$.

\begin{theorem}[Completeness]\label{th:Dinfr-completeness}
  Let $J\subseteq I$ be \emph{finite}. Let
  $M\in\Lamto$ and $\List x1n$ be pairwise distinct variables, such
  that
  \(
    \Tseq{\Family{x_i:m^j_i}i1n}{M}{b_j}
  \)
  in the $\Dinfr$ intersection type system for all $j\in J$.  Let
  $\List A1n$ and $B$ be
  formulas and let $\List u1n$ be almost injective functions such that
  $u_i:\Tdom{A_i}\to J=\Tdom B$. Assume also that, for all $j\in J$,
  one has $\Tfam B_j=b_j$ and $\Tfam{\Threloc{A_i}{u_i}}_j=m_i^j$ for
  $i=1,\dots,n$. Then the judgment
  \begin{align}\label{eq:th-untyped-compl-seq}
    \Jseq{\Threloc{A_1}{u_1},\dots,\Threloc{A_n}{u_n}}{B}
  \end{align}
  has a proof $\pi$ such that $\Tunderl\pi{\Vect x}\in\Qprojo M$.
\end{theorem}
Since the proof is very similar to that of
Theorem~\ref{th:simple-completeness}, we give it in an
Appendix.

\section{Concluding remarks}
The results presented in this paper show that, at least in
non-idempotent intersection types, the problem of knowing whether all
elements of a given family of intersection types $(a_j)_{j\in J}$ are
inhabited by a common $\lambda$-term can be reformulated logically: is
it true that one (or equivalently, any) of the indexed formulas $A$
such that $\Tdom A=J$ and $\forall j\in\ \Tfam A_j=a_j$ is provable in
$\LJ(I)$?  Such a strong connection between intersection and Indexed
Linear Logic was already mentioned in the introduction
of~\cite{BucciarelliEhrhard98}, but we never made it more explicit
until now.

To conclude we propose a typed $\lambda$-calculus \emph{à la Church}
to denote proofs of the $\LJ(I)$ system of
Section~\ref{sec:D-infty-relational}. The syntax of \emph{pre-terms}
is given by
\( s,t\dots \Bnfeq \Ivar xJ \Bnfor \Iabst xAus \Bnfor \Iapp st \)
where in $\Ivar xJ$, $x$ is a variable and $J\subseteq I$
and, in $\Iabst xAus$, $u$ is an almost injective function from
$\Tdom A$ to a set $J\subseteq I$. Given a pre-term $s$ and a variable
$x$, the \emph{domain of $x$ in $s$} is the subset $\Idomt xs$ of $I$
given by $\Idomt x{\Ivar xJ}=J$, $\Idomt x{\Ivar yJ}=\emptyset$ if
$y\not=x$, $\Idomt x{\Iabst yAus}=\Idomt xs$ (assuming of course
$y\not=x$) and $\Idomt x{\Iapp st}=\Idomt xs\cup\Idomt xt$. Then a
pre-term $s$ is a term if any subterm of $t$ which is of shape
$\Iapp{s_1}{s_2}$ satisfies
$\Idomt{x}{s_1}\cap\Idomt{x}{s_2}=\emptyset$ for all variable $x$. A
typing judgment is an expression
$\Tseq{\Family{x_i:\Threloc{A_i}{u_i}}i1n}{s}{B}$
where the $x_i$'s are pairwise distinct variables, $s$ is a term and
each $u_i$ is an almost injective function $\Tdom{A_i}\to\Tdom B$.
The following typing rules exactly mimic the logical rules of
$\LJ(I)$:
\begin{center}
  \AxiomC{$\Tdom A=\emptyset$}
  \UnaryInfC{$\Tseq{(\Family{x_i:\Threloc{A_i}{\Fempty\emptyset}}i1n)}{\Tinflam}{A}$}
  \DisplayProof
\end{center}

\begin{center}
  \AxiomC{$q\not=i\Implies\Tdom{A_i}=\emptyset$ and $u_i$ bijection}
  \UnaryInfC{$\Tseq{\Family{x_q:\Threloc{A_q}{u_q}}q1n}{\Ivar{x_i}{\Tdom{A_i}}}
    {\Treloc{u_i}{A_i}}$}
  \DisplayProof
  \quad
  \AxiomC{$\Tseq{\Family{x_i:\Threloc{A_i}{u_i}}i1n,x:\Threloc Au}{s}{B}$}
  \UnaryInfC{$\Tseq{\Family{x_i:\Threloc{A_i}{u_i}}i1n}{\Iabst xAus}
    {\Tfun AuB}$}
  \DisplayProof
\end{center}

\begin{center}
  \AxiomC{$\Tseq{\Family{x_i:\Threloc{\Trestr{A_i}{\Idomt{x_i}{s}}}
        {v_i}}i1n}{s}{\Tfun AuB}$}
  \AxiomC{$\Tseq{
      \Family{x_i:\Threloc{\Trestr{A_i}{\Idomt{x_i}{t}}}{w_i}}i1n}{t}{A}$}
  \BinaryInfC{$\Tseq{\Family{x_i:\Threloc{A_i}{v_i+ (u\Comp w_i)}}i1n}{\Iapp st}{B}$}
  \DisplayProof
\end{center}
The properties of this calculus, and more specifically of its
$\beta$-reduction, and its connections with the resource calculus
of~\cite{EhrhardRegnier04c} will be explored in further work.

Another major objective will be to better understand the meaning of
$\LJ(I)$ formulas, using ideas developed
in~\cite{BucciarelliEhrhard99} where a \emph{phase semantics} is
introduced and related to (non-uniform) coherence space semantics. In
the intuitionistic present setting, it is tempting to look for
Kripke-like interpretations with the hope of generalizing indexed
logic beyond the (perhaps too) specific relational setting we started
from.

\bibliographystyle{alpha}
\bibliography{newbiblio}

\section*{Appendix: proof of Theorem~\ref{th:Dinfr-completeness}}
\Beginproof
By induction on $M$.

Observe first that, if $J=\emptyset$, that is $\Tdom B=\emptyset$,
then $u_i=\Fempty\emptyset$ for $i=1,\dots,n$ and
therefore~\ref{eq:th-untyped-compl-seq} has a proof $\pi$ such that
$\Tunderl\pi{\Vect x}\Omegaeq\Tinflam$ 
by Lemma~\ref{lemme:empty-proofs}.

Assume that $M=\Tinflam$. Then we must have $J=\emptyset$ and so we
are in the situation described above.

Assume next that $M=x_i$ for some $i\in\Eset{1,\dots,n}$. Then we must
have $m_q=\Mset{}$ for $q\not=i$ and $m_i=\Mset b$. Therefore
$\Tdom{A_q}=\emptyset$ and $u_q=\Fempty J$ for $q\not=i$, $u_i$ is a
bijection $\Tdom{A_i}\to J$ and
$\forall k\in\Tdom{A_i}\ \Tfam{A_i}_k=b_{u_i(k)}$, in other words
$\Treloc{u_i}{A_i}\Tsim B$.
By Theorem~\ref{th:Dinfr-Tsim-implies} we know that the judgment
$\Jseq{\Threloc{(\Treloc{u_i}{A_i})}{\Id}}{B}$ is provable in $\LJ(I)$
with a proof $\rho$ such that $\Tunderl\rho x\in\Qproj x$ so that
using Lemma~\ref{lemma:substitution-2} and the axiom
$\Jseq{\Family{\Threloc{A_q}{u_q}}q1n}{\Treloc{u_i}{A_i}}$
we have a proof $\pi$ of
$\Jseq{\Family{\Threloc{A_q}{u_q}}q1n}{B}$ such that
$\Tunderl\pi{\Vect x}=\Subst{\Tunderl\rho
  x}{x_i}{x}=\Tunderl\rho{x_i}\in\Qproj{x_i}$.


Assume that $M=\Abs{x}{N}$ and that we have a family of deductions
(for $j\in J$) of
\(
  \Tseq{\Family{x_i:m^j_i}i1n}{M}
  {(m^j,c_j)}
\)
with $b_j=(m^j,c_j)$ and the premise of this conclusion in each of
these deductions is
\(
  \Tseq{\Family{x_1:m^j_i}i1n,x:m^j}{N}{c_j}
\).
By the observation at the beginning of this proof, we can assume
$J\not=\emptyset$ and since $\Tfam B_j=(m^j,c_j)$ for each $j\in J$,
we must have $B=\Tfunp CuD$ with $u:\Tdom C\to\Tdom D$ almost
injective, $\Tfam D_j=c_j$, $\Tfam{\Threloc Cu}_j=m^j$, for each
$j\in J$.
By inductive hypothesis we have a proof $\rho$ of
\(
  \Jseq{\Family{\Threloc{A_i}{u_i}}i1n,\Threloc Cu}{D}
\)
such that $\Tunderl\rho{\Vect x,x}\in\Qprojo N$ from which we obtain a
proof $\pi$ of
$\Jseq{\Family{\Threloc{A_i}{u_i}}i1n}{\Tfun CuD}$ such
that
$\Tunderl\pi{\Vect x}=\Abst x\sigma{\Tunderl\rho{\Vect x,x}}\in\Qprojo
M$ as expected.

Assume last that $M=\App NP$ and that we have a $J$-indexed family of
deductions
\(
  \Tseq{\Family{x_i:m^j_i}i1n}{M}{b_j}
\).
Let $\List A1n$, $\List u1n$ and $B$ be 
$\LJ(I)$ formulas and almost injective functions as in the statement
of the theorem.

Let $j\in J$. There is a finite set $L_j\subseteq I$ and multisets
$m_i^{j,0}$, $(m_i^{j,l})_{l\in L_j}$ such that we have deductions of
\(
  \Tseq{\Family{x_i:m^{j,0}_i}i1n}{N}
  {(\Mset{a^j_l\St l\in L_j},b_j)}
\)
and, for each $l\in L_j$, of
\(
  \Tseq{\Family{x_i:m^{j,l}_i}i1n}{P}{a^j_l}
\)
with
\begin{align}\label{eq:mset-hyp-decomp-u}
  m_i^j=m_i^{j,0}+\sum_{l\in L_j}m_i^{j,l}\,.
\end{align}
We assume the finite sets $L_j$ to be pairwise disjoint (this is
possible because $I$ is infinite) and we use $L$ for their union which
is finite since $J$ is finite. Let $u:L\to J$ be the function which
maps $l\in L$ to the unique $j$ such that $l\in L_j$, this function is
almost injective (actually, it is finite!). Let $A$ be a 
$\LL(J)$ formula such that $\Tunder A=\sigma$, $\Tdom A=L$ and
$\Tfam A_l=a_l^{u(l)}$; such a formula exists by
Proposition~\ref{prop:dinf-family-rep}.


Let $i\in\Eset{1,\dots,n}$. For each $j\in J$ we know that
\begin{align*}
  \Mset{\Tfam{A_i}_r\St r\in\Tdom{A_i}\text{ and }u_i(r)=j}=m_i^j
  =m_i^{j,0}+\sum_{l\in L_j}m_i^{j,l}
\end{align*}
and hence we can split the set $\Tdom{A_i}\cap\Funinv{u_i}(\Eset{j})$
into disjoint subsets $R_i^{j,0}$ and $(R_i^{j,l})_{l\in L_j}$ in such
a way that
\begin{align*}
  \Mset{\Tfam{A_i}_r\St r\in R_i^{j,0}}=m_i^{j,0}
  \text{\quad and\quad}
  \forall l\in L_j\ \Mset{\Tfam{A_i}_r\St r\in R_i^{j,l}}=m_i^{j,l}\,.
\end{align*}
We set $R_i^0=\Union_{j\in J}R_i^{j,0}$; observe that this is a
disjoint union because
$R_i^{j,0}\subseteq\Funinv{u_i}(\Eset{j})$. Similarly we define
$R_i^1=\Union_{l\in L}R_i^{u(l),l}$ which is a disjoint union for the
following reason: if $l,l'\in L$ satisfy $u(l)=u(l')=j$ then
$R_i^{j,l}$ and $R_i^{j,l'}$ have been chosen disjoint and if $u(l)=j$
and $u(l')=j'$ with $j\not=j'$ we have
$R_i^{j,l}\subseteq\Funinv{u_i}{\Eset j}$ and
$R_i^{j',l'}\subseteq\Funinv{u_i}{(\Eset{j'})}$.  Let $v_i:R_i^1\to L$
be defined by: $v_i(r)$ is the unique $l\in L$ such that
$r\in R_i^{u(l),l}$. Since each $R_i^{j,l}$ is finite the function
$v_i$ is almost injective (again, it is actually finite). Moreover
$u\Comp v_i=\Frestr{u_i}{R_i^1}$.


We use $u'_i$ for the restriction of $u_i$ to $R_i^0$ so that
$u'_i:R_i^0\to J$.  By inductive hypothesis
there is a proof $\mu$ of
\(
  \Jseq{\Family{\Threlocp{\Trestr{A_i}{R_i^0}}{u'_i}}i1n}{\Tfun{A}{u}{B}}
\)
such that $\Tunderl\mu{\Vect x}\in\Qprojo N$. Indeed
$\Mset{\Tfam{\Trestr{A_i}{R_i^0}}_r\St r\in R_i^0\text{ and
  }u'_i(r)=j}=m_i^{j,0}$ and
$\Tfam{\Tfun AuB}_j=(\Mset{a_l^j\St u(l)=j},b_j)$ for each $j\in J$.
By inductive hypothesis 
there is a proof $\rho$ of
\(
  \Jseq{\Family{\Threlocp{\Trestr{A_i}{R_i^1}}{v_i}}i1n}{A}
\)
such that $\Tunderl\rho{\Vect x}\in\Qprojo P$. Indeed for each
$l\in L=\Tdom A$ we have
\[
  \Mset{\Tfam{\Trestr{A_i}{R_i^1}}_r\St v_i(r)=l}=m_i^{j,l}
\]
and $\Tfam A_l=a_l^{j}$ where $j=u(l)$. By a $\Implies$-elimination
rule we get a proof $\pi$ of
\( \Jseq{\Family{\Threloc{A_i}{u_i}}i1n}{B} \) such that
$\Tunderl\pi{\Vect x}=\App{\Tunderl\mu{\Vect x}}{\Tunderl\rho{\Vect
    x}}\in\Qprojo M$.
\Endproof

\label{lastpage}

\end{document}

%% file: local.tex
\newcommand\CMLLPAR{
\usepackage{cmll}
\newcommand\IPar{\mathord{\parr}}
}

\CMLLPAR

%% file: notation.tex
\renewcommand\paragraph{\subsubsection}

\newcommand{\proofitem}[1]{\paragraph*{\mdseries\textit{#1}}}

\newcommand{\Beginproof}{\proofitem{Proof.}}
\newcommand{\Endproof}{
  \ifmmode 
  \else \leavevmode\unskip\penalty9999 \hbox{}\nobreak\hfill
  \fi
  \quad\hbox{$\Box$}
  \par\medskip}

\renewcommand{\phi}{\varphi}
\renewcommand\epsilon{\varepsilon}

\newcommand{\Implies}{\Rightarrow}

\newcommand{\St}{\mid}

\newcommand\Seqempty{\langle\rangle}

\newcommand\cH{\mathcal{H}}

\newcommand\Fini{{\mathrm{fin}}}

\newcommand\Part[1]{{\mathcal P}({#1})}

\newcommand\Union{\bigcup}

\newcommand{\Linarrow}{\multimap}

\newcommand\Mset[1]{[\,{#1}\,]}
\newcommand\Eset[1]{\{#1\}}

\newcommand\IWith{\mathrel{\&}}
\newcommand\With[2]{{#1}\IWith{#2}}

\newcommand\Bwith{\mathop{\&}}

\newcommand\LL{\textsf{LL}}
\newcommand\LJ{\textsf{LJ}}

\newcommand\Fempty[1]{0_{#1}}

\newcommand\Card[1]{\#{#1}}

\newcommand\Locun[1]{1^J}

\newcommand\Isom\simeq

\newcommand\Comp{\mathrel\circ}

\newcommand\Funinv[1]{{#1}^{-1}}

\newcommand\Limpl[2]{{#1}\Linarrow{#2}}

\newcommand\Nat{{\mathbb{N}}}

\newcommand\App[2]{\left({#1}\right){#2}}
\newcommand\Abs[2]{\lambda{#1}\,{#2}}
\newcommand\Abst[3]{\lambda{#1}^{#2}\,{#3}}

\newcommand\List[3]{#1_{#2},\dots,#1_{#3}}

\newcommand\Subst[3]{{#1}\left[{#2}/{#3}\right]}

\newcommand\Mfin[1]{\mathcal M_\Fini({#1})}

\newcommand\Ev{\operatorname{\mathsf{Ev}}}

\newcommand\RELK{\operatorname{\mathbf{Rel}_\oc}}

\newcommand\Rel[1]{\mathrel{#1}}

\newcommand\Redst[1]{\mathop{\mathsf{Red}}}

\newcommand\Varset{\mathcal{V}}

\newcommand\Tuple[1]{\langle{#1}\rangle}

\newcommand\Msetofsubst[1]{\bar F}

\newcommand\Retri\zeta
\newcommand\Retrp\rho

\newcommand\Impl[2]{{#1}\Rightarrow{#2}}

\newcommand\Tsem[1]{[{#1}]}

\newcommand\Tnat\iota

\newcommand\Loop\Omega

\newcommand\Tseq[3]{{#1}\vdash{#2}:{#3}}

\newcommand\Timpl\Impl
\newcommand\Timplp[2]{(\Timpl{#1}{#2})}
\newcommand\Simpl\Impl

\newcommand\Curry[1]{\operatorname{\mathsf{Cur}}({#1})}

\newcommand\Emptyseq{\Seqempty}

\newcommand\Id{\operatorname{\mathsf{Id}}}

\newcommand\Proj[1]{\operatorname{\mathsf{pr}}_{#1}}

\newcommand\Relincl\eta
\newcommand\Relrestr\rho

\newcommand\Vect[1]{\overrightarrow{#1}}

\newcommand\Eta{\leadsto_\eta}
\newcommand\Inflam{\leadsto_\omega}

\newcommand\Etaeq{\sim_{\eta}}
\newcommand\Omegaeq{\sim_{\omega}}
\newcommand\Beoeq{\sim_{\beta\eta\omega}}
\newcommand\Etaoeq{\sim_{\eta\omega}}

\newcommand\Eqref[1]{(\ref{#1})}

\newcommand\Treelocl[1]{\Emptyseq}

\newcommand\Tunder[1]{\underline{#1}}
\newcommand\Tunderl[2]{\underline{#1}_{#2}}
\newcommand\Tdom[1]{\operatorname{\mathsf d}(#1)}
\newcommand\Tfam[1]{\langle #1\rangle}
\newcommand\Tcst[2]{#1[#2]}
\newcommand\Tfun[3]{#1\Rightarrow_{#2}#3}
\newcommand\Tfunp[3]{(#1\Rightarrow_{#2}#3)}
\newcommand\Trestr[2]{#1\mathord\restriction_{#2}}
\newcommand\Frestr[2]{#1\mathord\restriction_{#2}}
\newcommand\Funrestr[2]{#1\restriction_{#2}}
\newcommand\Treloc[2]{{#1}_*(#2)}
\newcommand\Threloc[2]{#1^{#2}}
\newcommand\Threlocp[2]{(#1)^{#2}}

\newcommand\Jseq[2]{#1\vdash #2}

\newcommand\Seqext[2]{(#1)\setminus{#2}}

\newcommand\Tsim{\sim}

\newcommand\Dinfr{\mathsf R_\infty}
\newcommand\Dinfrsize{\operatorname{\mathsf{sz}}}

\newcommand\Estack{\mathsf e}
\newcommand\Estackf[1]{\epsilon_{#1}}

\newcommand\Otype{\mathsf o}

\newcommand\Tinflam{\Omega}

\newcommand\Lamt{\Lambda}
\newcommand\Lamto{\Lambda_\Tinflam}

\newcommand\Qproj[1]{\cH(#1)}
\newcommand\Qprojo[1]{\cH_\Omega(#1)}
\newcommand\Appsep{\,}

\newcommand\Family[4]{(#1)_{#2=#3}^{#4}}

\newcommand\Csucc[2]{\operatorname{\mathsf s}(#1,#2)}

\newcommand\Bnfeq{\Rel{\mathord:\mathord=}}
\newcommand\Bnfor{\mid}

\newcommand\Ivar[2]{#1[#2]}
\newcommand\Iabst[4]{\lambda #1:#2^{#3}\,#4}
\newcommand\Iapp[2]{\App{#1}{#2}}

\newcommand\Idomt[2]{\operatorname{\mathsf{dom}}(#1,#2)}



%% file: tikznotation.tex
\usepackage{tikz-inet}

\tikzset{inetwire/.style={line width=0.15ex,rounded corners=2pt}}